\documentclass[prd,preprint,showpacs]{revtex4-1}
\usepackage{amsmath}
\usepackage{latexsym}
\usepackage{amsfonts}
\usepackage{graphicx}
\usepackage{mathrsfs}
\usepackage{CJK}
\usepackage{longtable}

\usepackage{hyperref}
\hypersetup{
  colorlinks = true,
  urlcolor = blue,
  linkcolor = blue,
  citecolor = green,
  filecolor = magenta,
}

\newcommand{\ud}{\mathrm{d}}

\newcommand{\script}{\mathscr}

\begin{document}
\begin{CJK*}{GB}{gbsn}

\title{Strong Equivalence Principle and Gravitational Wave Polarizations in Horndeski Theory}
\author{Shaoqi Hou}
\email{shou1397@hust.edu.cn}
\affiliation{School of Physics, Huazhong University of Science and Technology, Wuhan, Hubei 430074, China}
\author{Yungui Gong}
\email{yggong@hust.edu.cn}
\affiliation{School of Physics, Huazhong University of Science and Technology, Wuhan, Hubei 430074, China}
\date{\today}

\begin{abstract}
The relative acceleration between two nearby particles moving along accelerated trajectories is studied, which generalizes the geodesic deviation equation.
The polarization content of the gravitational wave in Horndeski theory is investigated by examining the relative acceleration between two self-gravitating particles.
It is found out that the longitudinal polarization exists no matter whether the scalar field is massive or not.
It would be still very difficult to detect the enhanced longitudinal polarization  with the interferometer, as the violation of the strong equivalence principle of mirrors used by interferometers is extremely small.
However, the pulsar timing array is promised  relatively easily to detect the effect of the violation as  neutron stars have large self-energy.
The advantage of using this method to test the violation of the strong equivalence principle is that neutron stars are not required to be present in the binary systems.
\end{abstract}

\maketitle
\end{CJK*}

\section{Introduction}

Soon after the birth of General Relativity (GR), several alternative theories of gravity were proposed.
The discovery of the accelerated expansion of the Universe \cite{Perlmutter:1998np,Riess:1998cb} revives the pursuit of these alternatives because the extra fields might account for the dark energy.
Since Sep.~14th, 2015, LIGO/Virgo collaborations have detected ten gravitational wave (GW) events \cite{Abbott:2016blz,Abbott:2016nmj,Abbott:2017vtc,Abbott:2017oio,TheLIGOScientific:2017qsa,Abbott:2017gyy,LIGOScientific:2018mvr}.
This opens a new era of probing the nature of gravity in the highly dynamical, strong-field regime.
Due to the extra fields, alternatives to GR generally predict that there are extra GW polarizations in addition to the plus and cross ones in GR.
So the detection of the polarization content is very essential to test whether GR is \emph{the} theory of gravity.
In GW170814,  the polarization content of GWs was measured for the first time, and the pure tensor polarizations were favored against pure vector and pure scalar polarizations \cite{Abbott:2017oio}.
Similar results were reached in the recent analysis on GW170817 \cite{Abbott:2018lct}.
More interferometers are needed to finally pin down the polarization content.
Other detection methods might also determine the polarizations of GWs such as pulsar timing arrays (PTAs) \cite{Kramer:2013kea,Hobbs:2009yy,McLaughlin:2013ira,Hobbs:2013aka}.

Alternative metric theories of gravity may not only introduce extra GW polarizations, but also violate the strong equivalence principle (SEP) \cite{Will:1993ns} \footnote{Nordstrom's scalar theory of gravity satisfies SEP, but is not a metric theory, not to mention that it has been excluded by the observations \cite{Deruelle:2011wu}.}.
The violation of strong equivalence principle (vSEP) is due to the extra degrees of freedom,
which indirectly interact with the matter fields via the metric tensor.
This indirect interaction modifies the self-gravitating energy of the objects and leads to vSEP \cite{Barausse:2015wia}.
The self-gravitating objects no longer move along geodesics, even if there is only gravity acting on them, and the relative acceleration between the nearby objects does not follow the geodesic deviation equation.
In the usual approach, one assumes that the test particles, such as the mirrors in the aLIGO, move along geodesics, so their relative acceleration is given by the geodesic deviation equation.
Since the polarization content of GWs is determined by examining the relative acceleration,
the departure from the geodesic motion might effectively result in different polarization contents, which can be detected by PTAs.
Thus, the main topic of this work is to investigate the effects of vSEP on the polarization content of GWs and the observation of PTAs.

To be more specific, the focus is on the vSEP in the scalar-tensor theory, which is the simplest alternative metric theory of gravity.
The scalar-tensor theory contains one scalar field $\phi$ besides the metric tensor field $g_{\mu\nu}$ to mediate the gravitational interaction.
Because of the trivial transformation of the scalar field under the diffeomorphism, there are a plethora of scalar-tensor theories,
such as Brans-Dicke theory \cite{Brans:1961sx}, Einstein-dilaton-Gauss-Bonnet gravity (EdGB) \cite{Kanti:1995vq} and $f(R)$ gravity \cite{Buchdahl:1983zz,OHanlon:1972xqa,Teyssandier:1983zz}.
In 1974, Horndeski constructed the most general scalar-tensor theory \cite{Horndeski:1974wa}.
Its action contains higher derivatives of $\phi$ and $g_{\mu\nu}$, but still gives rise to at most the second order differential field equations.
So the Ostrogradsky instability is absent in this theory \cite{Ostrogradsky:1850fid}.
In fact, Horndeski theory includes previously mentioned theories as its subclasses.
In this work,  the vSEP in Horndeski theory will be studied.

Among the effects of  vSEP, Nordtvedt effect is well-known for a long time \cite{Nordtvedt:1968qr,Nordtvedt:1968qs}, and happens in the near zone of the source of the gravitational field. It leads to observable effects.
For example, the Moon's orbit around the Earth will be polarized when they are moving in the gravitational field generated by the Sun \cite{1982RPPh...45..631N,Will:2014kxa}.
The polarization of the Moon's orbit has been constrained by the lunar laser ranging experiments \cite{Alsing:2011er}, which gave the Nordtvedt parameter \cite{2010A&A...522L...5H}
\begin{equation}\label{nordpar}
  \eta_\text{N}=(0.6\pm5.2)\times10^{-4},
\end{equation}
which measures vSEP in the following way,
\begin{equation}\label{eq-def-nf}
  \frac{m_g}{m_i}=1+\eta_\text{N}\varepsilon_\text{grav.}+O(\varepsilon_\text{grav.}^2),
\end{equation}
with $m_g$ and $m_i$ the gravitational and the inertial masses, and $\varepsilon_\text{grav.}$ the ratio of the gravitational binding energy to the inertial energy.
A similar polarization of the orbit of the millisecond pulsar-white dwarf (MSP-WD) system also happens due to the gravitational field of the Milky Way \cite{Damour:1991rq,Freire:2012nb}.
In contrast with the Moon and the Earth, pulsars have large gravitational binding energies, so the observation of the orbit polarization of MSP-WD systems set constraints on vSEP in the strong field regime, which was discussed in Ref.~\cite{Freire:2012nb}.
The observation of a triple pulsar PSR J0337+1715 was used to  set $\Delta=(-1.09\pm0.74)\times10^{-6}$ \cite{Archibald:2018oxs}.
The vSEP also leads to the dipole gravitational radiation, and the variation of Newton's constant $G$ \cite{Stairs:2003eg}.
The dipole gravitational radiation for Horndeski theory has been studied in Ref.~\cite{Hou:2017cjy}, and constraints on this theory were obtained.
The pulsar timing observation of the binary system J1713+0747 has leads to $\dot G/G=(-0.1\pm0.9)\times10^{-12}\text{ yr}^{-1}$ and $|\Delta|<0.002$ \cite{Zhu:2018etc}.

As discussed above, none of the previous limits on vSEP was obtained directly using the GW.
So probing vSEP by measuring the GW polarizations provides a novel way to test GR in the high speed and dynamical regime.
It will become clear that although the vSEP will effectively enhance the longitudinal polarization, it is still very difficult for aLIGO to detect the effects of the longitudinal polarization, as the vSEP  by the mirror is extremely weak.
In contrast, neutron stars are compact objects with non-negligible self-gravitating energies.
The vSEP by neutron stars is strong enough that the stochastic GW background will affect their motions, which is reflected in the cross-correlation function for PTAs \cite{Jenet:2005pv,2008ApJ...685.1304L,Lee:2010cg,Lee:2014awa}.
By measuring the cross-correlation function, it is probably easier to detect the presence of vSEP.
For this purpose, one only has to observe the change in the arriving time of radial pulses from neutron stars without requiring the neutron stars be in binary systems.

This work is organized as follows.
Section \ref{sec-geodev} reviews the derivation of the geodesic deviation equation, and a generalized deviation equation for accelerated particles is discussed in Section \ref{sec-nongeodev}.
Section \ref{sec-horn} derives the motion of a self-gravitating object in presence of GWs in Horndeski theory.
The polarization content of GWs in Horndeski theory is revisited by taking the vSEP into account in Section \ref{sec-pols}.
The generalized deviation equation is computed to reveal the polarization content of GWs.
Section \ref{sec-pta} calculates the cross-correlation function for PTAs due to GWs.
Finally, Section \ref{sec-con} briefly summarizes this work.
Penrose's abstract index notation  is used \cite{Penrose1984v1}.
The units is chosen such that the speed of light $c=1$ in vacuum.

\section{Geodesic Deviation Equation}\label{sec-geodev}

This section serves to review the idea to derive the geodesic deviation equation following Ref.~\cite{Wald:1984rg}. In the next section, the derivation will be generalized to accelerated objects straightforwardly.

Let $\gamma_s(t)$ represent a geodesic congruence, in which each geodesic is parameterized by $t$ and labeled by $s$. Define the following tangent vector fields,
\begin{equation}\label{deftang}
  T^a=\left(\frac{\partial}{\partial t}\right)^a,\quad S^a=\left(\frac{\partial}{\partial s}\right)^a.
\end{equation}
$S^a$ is called the deviation vector. Their commutator vanishes,
\begin{equation}\label{eq-com}
  T^b\nabla_b S^a=S^b\nabla_bT^a.
\end{equation}
With a suitable parametrization, one requires that $T^b\nabla_b T^a=0$ so that $t$ is an affine parameter.
Note that it is not necessary to set $T^aT_a=-1$ for the following discussion.
Whenever desired, one can always reparameterize to normalize it.
It is now ready to derive the geodesic deviation equation,
\begin{equation}\label{dergeodev}
  A^a_\text{rel}=T^c\nabla_c(T^b\nabla_bS^a)=-R_{cbd}{}^aT^cS^bT^d,
\end{equation}
using Eq.~\eqref{eq-com}.
For details of derivation, please refer to Ref.~\cite{Wald:1984rg}.

The deviation vector $S^a$ is not unique.
A new parametrization of the geodesics,
\begin{equation}\label{eq-repar}
t\rightarrow t'=\alpha(s)t+\beta(s),
\end{equation}
results in the change in $S^a$ by a multiple of $T^a$,
\begin{equation}\label{ntang}
  T'^a=\frac{T^a}{\alpha(s)},\quad S'^a=S^a+\frac{\ud}{\ud s'}\left(\frac{t'-\beta(s')}{\alpha(s')}\right)T^a.
\end{equation}
Therefore, there is a gauge freedom in choosing the deviation vector field $S^a$. This gauge freedom will be used frequently below to simplify the analysis.

Firstly, there is a parametrization such that $T_aT^a$ is a constant along the coordinate lines of the constant $t$, i.e., the integral curves of $S^a$. In fact, one knows that,
\begin{equation}\label{cons}
  S^b\nabla_b(T_aT^a)=2T_aS^b\nabla_bT^a=2T_aT^b\nabla_bS^a,
\end{equation}
and under the reparameterization \eqref{eq-repar}, one gets
\begin{equation}\label{cons1}
  S'^b\nabla_b(T'_aT'^a)=\frac{2}{\alpha(s)^2}T_aT^b\nabla_b\left[S^a+\frac{\partial}{\partial s}\left(\frac{t'-\beta(s)}{\alpha(s)}\right)T^a\right],
\end{equation}
so it is always possible to choose a parametrization to achieve that $S'^b\nabla_b(T'_aT'^a)=0$.
Physically, this means that all geodesics are parameterized by the ``same" affine parameter $t'$.
Secondly, under the above parametrization, the inner product $T^aS_a$ can be made constant along the geodesics,
\begin{equation}\label{cont}
\begin{split}
  T^b\nabla_b(T^aS_a) =&T_aT^b\nabla_bS^a=T_aS^b\nabla_bT^a\\
  =&\frac{1}{2}S^b\nabla_b(T^aT_a)  =0.
  \end{split}
\end{equation}
An initial choice of $T^aS_a=0$ will be preserved along the $t$ coordinate line, so that $S^a$ is always a spatial vector field for an observer with 4-velocity $u^a=T^a/\sqrt{-T_bT^b}$ along its trajectory.

From the derivation, one should be aware that the geodesic deviation equation \eqref{dergeodev} is independent of the gauge choices made above, which only serves to make sure $S^a$ is always a spatial vector relative to an observer with $u^a$.
In this way, there is no deviation in the time coordinate, that is, no time dilatation.
This is because one concerns the change in the spatial distance between two nearby particles measured by either one of them.

\section{Non-geodesic Deviation Equation}\label{sec-nongeodev}

When particles are accelerated, they are not moving on geodesics. This happens when there are forces acting on these particles.
This also happens for self-gravitating particles in the modified gravity theories, such as the scalar-tensor theory. Suppose a bunch of particles are accelerated and therefore, their velocities satisfy the following relations,
\begin{equation}\label{non-geo}
  T^b\nabla_bT^a=A^a,
\end{equation}
with $A^a $ the 4-acceleration and not proportional to $T^a$.
In the following, $T^a$ is assumed to be some arbitrary timelike vector field which is not necessarily the 4-velocity of some particle.
In this general discussion, the only assumption is that $T^a$ satisfies Eq.~(\ref{non-geo}).
Now, the non-geodesic deviation equation can be derived similarly,
\begin{equation}\label{nongeodev}
     A^a_\mathrm{rel} =-R_{cbd}{}^aT^cS^bT^d+S^b\nabla_bA^a.
\end{equation}
Again, the derivation of this result does not reply on the gauge fixing made similarly in the previous section or the one to be discussed below.
Compared with Eq.~(\ref{dergeodev}), there is one extra term, which is due to the fact that the trajectories are no longer geodesics.
This equation and a more general one were derived in Ref.~\cite{1983JMP....24..883S} using the definitions of curvature and torsion.
The authors did not discuss the suitable gauge for extracting physical results which will be presented below.

If $T^aA_a\ne0$, one can reparameterize the integral curves of $T^a$ to make it vanish. Indeed, a reparameterization $t\rightarrow t'=\kappa(t)$ leads to
\begin{equation}\label{repnon-geo}
  A'^a =T'^b\nabla_bT'^a =\frac{A^a }{\dot\kappa^2}-\frac{\ddot\kappa}{\dot\kappa^3}T^a,
\end{equation}
where dot denotes the derivative with respect to $t$. So one can always find a new parametrization which annihilates $T'^aA'_a$, that is,
\begin{equation}\label{newpar}
  \kappa(t)=\alpha\int\exp\left(\frac{A^a T_a}{T^bT_b}t\right)\ud t+\beta,
\end{equation}
with $\alpha,\,\beta$ integration constants. From now on, $T^aA_a=0$ is assumed which implies that
\begin{equation}\label{consttt}
  T^b\nabla_b(T^aT_a)=0.
\end{equation}
So although $t$ may not be the proper time $\tau$, it is a linear function of $\tau$.
A further reparameterization $t'=\alpha't+\beta'$ does not change the above relation.

Now, pick a congruence of these trajectories $\sigma_s(t)$.
So as in the previous section, $\sigma_s(t)$'s also lie on a 2-dimensional surface $\Sigma$ parameterized by $(t,s)$.
There also exists the similar gauge freedom to that discussed in Section \ref{sec-geodev}, except that $A^a $ depends on the gauge choice.
For example, a  reparametrization $t\rightarrow t'=\alpha(s)t+\beta(s)$  results in changes in $S^a$ (given by Eq.~\eqref{ntang}) and $A^a $, i.e., $A^a \rightarrow A^a /\alpha^2(s)$.

With this gauge freedom, one also chooses a suitable gauge such that $T^aS_a$ remains constant along each trajectory. In fact, it can be shown that
\begin{equation}\label{constnon}
  T^b\nabla_b(T^aS_a)=S_aA^a +\frac{1}{2}S^b\nabla_b(T^aT_a).
\end{equation}
One requires that $T^aS_a=0$ along the integral curves of $T^a$, i.e., $T^b\nabla_b(T^aS_a)=0$. This implies that
\begin{equation}\label{constnon-c}
  S^b\nabla_b(T^aT_a)=-2S_aA^a .
\end{equation}
This expression means that if the trajectory $\sigma_0(t)$ is parameterized by the proper time $t=\tau$, a nearby trajectory $\sigma_s(t)$ with $s\ne0$ will not be parameterized by its proper time, in general.
It is necessary to choose this particular gauge as $S^a$ can be viewed as a spatial vector field relative to $T^a$ as long as $T^a$ can be interpreted as the 4-velocity of  an observer.

\subsection{Fermi normal coordinates}

In this subsection, the relative acceleration will be expressed in the Fermi normal coordinate system of the observer $\sigma_0(\tau)$ with $\tau$ the proper time.
Let the observer $\sigma_0(\tau)$ carry a pseudo-orthonomal tetrad $\{(e_{\hat 0})^a=u^a,(e_{\hat 1})^a,(e_{\hat 2})^a,(e_{\hat 3})^a\}$, which satisfies $g_{ab}(e_{\hat\mu})^a(e_{\hat\nu})^b=\eta_{\hat \mu\hat\nu}$ and is Fermi-Walker transported along $\sigma_0(\tau)$.  The observer $\sigma_0(\tau)$ will measure the deviation in its own proper reference frame, in which the metric takes the following form \cite{mtw},
\begin{equation}\label{leng}
  \ud s^2=-(1+2A_{\hat j}x^{\hat j})\ud \tau^2+\delta_{\hat j\hat k}\ud x^{\hat j}\ud x^{\hat k}+O(|x^{\hat j}|^2),
\end{equation}
where $j,\,k=1,2,3$ and the acceleration of $\sigma_0(\tau)$ has no time component ($A^{\hat  0}=-u_aA^a =0$).
Similarly, $S^a=S^{\hat j}(e_{\hat j})^a$, so the relative acceleration has the following spatial components
\begin{equation}\label{eq-rel-fn-spl}
\begin{split}
  A_\text{rel}^{\hat j}=&-R_{\hat 0\hat k\hat 0}{}^{\hat j}S^{\hat k}+S^{\hat k}\nabla_{\hat k}A^{\hat  j}\\
  =&-R_{\hat 0\hat k\hat 0}{}^{\hat j}S^{\hat k}+S^{\hat k}\partial_{\hat k}A^{\hat  j},
  \end{split}
\end{equation}
since the only nonvanishing components of the Christoffel symbol are
\begin{equation}\label{non0ch}
  \Gamma^{\hat0}{}_{\hat 0\hat j}=\Gamma^{\hat j}{}_{\hat0\hat0}=A_{\hat j}.
\end{equation}
The relative acceleration can also be expanded as
\begin{equation}\label{accexp}
  A^{\hat j}_\mathrm{rel}=u^{\hat\mu}\nabla_{\hat\mu}(u^{\hat\nu}\nabla_{\hat\nu}S^{\hat j})
  =\frac{\ud^2S^{\hat j}}{\ud\tau^2}+A^{\hat  j}A_{\hat k}S^{\hat k}.
\end{equation}
Therefore, one gets
\begin{equation}\label{eq-fn-sjdd}
  \frac{\ud^2S^{\hat j}}{\ud\tau^2}=-R_{\hat 0\hat k\hat 0}{}^{\hat j}S^{\hat k}+S^{\hat k}\partial_{\hat k}A^{\hat  j}-A^{\hat  j}A_{\hat k}S^{\hat k}.
\end{equation}
Similar expression was also found in Ref.~\cite{Hawking:1973uf}.
Due to the requirement $T^b\nabla_b(S^aT_a)=0$, one knows that $\ud^2S^{\hat 0}/\ud\tau^2=T^c\nabla_c[T^b\nabla_b(S^aT_a)]=0$, so $S^{\hat 0}=0$ is really preserved while $S^a$ is propagated along the integral curves of $T^a$.
Whenever the observer $\sigma_0(\tau)$ is moving on a geodesic, $A^a=0$, then Eq.~\eqref{eq-fn-sjdd} becomes the usual geodesic deviation equation used to analyze the polarizations of GWs \cite{mtw}.

\section{The Trajectory of a Self-gravitating Object in Horndeski Theory}\label{sec-horn}

The most general scalar-tensor theory with second order equations of motion is the Horndeski theory \cite{Horndeski:1974wa}, whose action is given by \cite{Kobayashi:2011nu},
\begin{equation}\label{acth}
  S=\int\ud^4x\sqrt{-g}(\mathscr L_2+\mathscr L_3+\mathscr L_4+\mathscr L_5)+S_m[\psi_m,g_{\mu\nu}],
\end{equation}
where $S_m[\psi_m,g_{\mu\nu}]$ is the action for the matter field $\psi_m$, and it is assumed that $\psi_m$ non-minimally couples with the metric only.
The individual terms in the integrand are
\begin{gather}
  \mathscr L_2 = K(\phi,X),\\
  \script L_3=-G_3(\phi,X)\Box\phi, \\
  \script L_4 = G_4(\phi,X)R+G_{4X}[(\Box\phi)^2-(\phi_{;\mu\nu})^2], \\
  \script L_5 = G_5(\phi,X)G_{\mu\nu}\phi^{;\mu\nu}-\frac{G_{5X}}{6}[(\Box\phi)^3-3(\Box\phi)(\phi_{;\mu\nu})^2+2(\phi_{;\mu\nu})^3].
\end{gather}
In these expressions, $X=-\phi_{;\mu}\phi^{;\mu}/2$ with $\phi_{;\mu}=\nabla_\mu\phi$, $\phi_{;\mu\nu}=\nabla_\nu\nabla_\mu\phi$, $\Box\phi=g^{\mu\nu}\phi_{;\mu\nu}$, $(\phi_{;\mu\nu})^2=\phi_{;\mu\nu}\phi^{;\mu\nu}$ and $(\phi_{;\mu\nu})^3=\phi_{;\mu\nu}\phi^{;\mu\rho}\phi^{;\nu}_{;\rho}$ for simplicity.
$K, G_3, G_4, G_5$ are arbitrary analytic functions of $\phi$ and $X$, and $G_{iX}=\partial_XG_{i}, i=3,4,5$.
For any binary function $f(\phi,X)$, define the following symbol
\begin{equation}\label{defsym}
  f_{(m,n)}=\frac{\partial^{m+n}f(\phi,X)}{\partial \phi^m\partial X^n}\Big|_{\phi=\phi_0,X=0},
\end{equation}
where $\phi_0$ is a constant value for the scalar field evaluated at infinity.
Varying the action \eqref{acth} with respect to $g_{\mu\nu}$ and $\phi$ gives rise to the equations of motion, which are too complicated to write down. Please refer to Refs~\cite{Kobayashi:2011nu,Gao:2011mz}.

There have been experimental constraints on Horndeski theory.
Ref.~\cite{Hou:2017cjy} discussed the bounds on it from some solar system tests and the observations on pulsars.
GW170817 and its electromagnetic counterpart GRB 170817A together set a strong constraint on the speed of GWs \cite{TheLIGOScientific:2017qsa,Monitor:2017mdv}.
Based on this result, the Lagrangian takes a simpler form \cite{Lombriser:2015sxa,Lombriser:2016yzn,Baker:2017hug,Creminelli:2017sry,Sakstein:2017xjx,Ezquiaga:2017ekz,Langlois:2017dyl,Sakstein:2017xjx,Gong:2017kim},
\begin{equation}\label{eq-acth-c}
  \mathscr L=K(\phi,X)-G_3(\phi,X)\Box\phi+G_4(\phi)R.
\end{equation}
Although Horndeski theory is highly constrained, we will still work with the original theory in the following discussion.

In this theory, WEP is respected due to the non-minimal coupling between $\psi_m$ and $g_{\mu\nu}$.
However, due to the indirect interaction between $\psi_m$ and $\phi$ mediated by $g_{\mu\nu}$ via the equations of motion, SEP is violated.
In fact, calculations have shown that the effective gravitational ``constant" actually depends on $\phi$ \cite{Hohmann:2015kra}.
Therefore, the gravitational binding energy of a compact object, viewed as a system of point particles, will also depend on the local value of $\phi$.
Because of the mass-energy equivalence $E=m$, the mass of the compact object, i.e., the total mass of the system of point particles, also depends on $\phi$.
This would affect the motion of the compact object.
Following Eardley's suggestion, the matter action  can be described by \cite{1975ApJ...196L..59E}
\begin{equation}\label{actsg}
  S_m=-\int m(\phi(x^\rho))\sqrt{-g_{\mu\nu}(x^\rho)\dot x^\mu\dot x^\nu}\ud\lambda,
\end{equation}
with $\dot x^\mu=\ud x^\mu/\ud\lambda$, when the compact object can be treated as a self-gravitating particle.
In this action, $\phi$ and $g_{\mu\nu}$ also depend on the trajectory.
In this treatment, the spin and the multipole moment structure are ignored.
To obtain the equation of motion, one applies Euler-Lagrange equation and at the same time, assumes that the parameter $\lambda$ parameterizes the trajectory such that $g_{\mu\nu}\dot x^\mu\dot x^\nu$ is a constant along the trajectory.
Usually, one parameterizes particle trajectories with the proper time $\tau$.
This is not necessary, as one can always reparameterize.
A generic parametrization is convenient for the following discussion.

The Euler-Lagrange equation reads,
\begin{equation}\label{elnongeo}
  A^a=u^b\nabla_bu^a
  =-\frac{\ud\ln m}{\ud\ln \phi}(-g^{ab}u_cu^c+u^au^b)\nabla_b\ln\phi,
\end{equation}
where $u^a=(\partial/\partial\lambda)^a$.
Therefore, the self-gravitating particle no longer moves on a geodesic.
The failure of its trajectory being a geodesic is described by  $\frac{\ud\ln m}{\ud\ln \phi}$, which is called the "sensitivity".
One can check  that $u_au^b\nabla_bu^a=0$, which is consistent with the parametrization.
This means that the 4-acceleration of the particle is a spatial vector with respect to $u^a$.
If one  chooses the proper time $\tau$ to parameterize the trajectory, the above expression gets simplified,
\begin{equation}\label{elnongeos}
  A^a=-\frac{\ud\ln m}{\ud\ln \phi}(\delta^a_b+u^au_b)\nabla^b\ln\phi,
\end{equation}
where $\delta^a_b+u^au_b$ is actually the projection operator for $u^a$.
Therefore, a self-gravitating object moves along an accelerated trajectory when only gravity acts on it, and its acceleration is due to the gradient in the scalar field $\phi$.

Now, consider two infinitesimally nearby self-gravitating particles, one of which travels along $\sigma_0(\lambda)$.
The deviation vector connecting $\sigma_0(\lambda)$ to its nearby company is $S^a$.
It is useful to parameterize $\sigma_0(\lambda)$ by its proper time $\tau$ so that $u^a$ is a unit timelike vector associated with an observer.
The relative acceleration is thus given by
\begin{equation}\label{relaccst}
\begin{split}
  A^a_\mathrm{rel}=&-R_{cbd}{}^au^cS^bu^d\\
  &-S^b\nabla_b\left[\frac{\ud\ln m}{\ud\ln \phi}(-g^{ac}u_du^d+u^au^c)\nabla_c\ln\phi\right].
  \end{split}
\end{equation}
Note that the right hand side is evaluated at $\sigma_0(\tau)$.
The deviation vector $S^a$ should satisfy
\begin{gather}
u^b\nabla_bS^a=S^b\nabla_bu^a,\label{gcons-1}\\
  S^b\nabla_b(u^au_a)=2\frac{\ud\ln m}{\ud\ln \phi}S^a\nabla_a\ln\phi,\label{gcons-2}
\end{gather}
according to Eq.~(\ref{constnon-c}), which explains why $u_du^d$ inside of the brackets of Eq.~(\ref{relaccst}) is not set to $-1$.
The relative acceleration can be expressed entirely in terms of $u^a$ of the particle $\sigma_0(\tau)$ by expanding the brackets and using Eq.~(\ref{gcons-1}) together with Eq.~(\ref{gcons-2}),
\begin{equation}\label{relaccst1}
\begin{split}
  A^a_\mathrm{rel}=&-R_{cbd}{}^au^cS^bu^d
  -(g^{ac}+u^au^c)S^b\nabla_b\left(\frac{\ud\ln m}{\ud\ln\phi}\nabla_c\ln\phi\right)\\
  &-\frac{\ud\ln m}{\ud\ln\phi}(\nabla_c\ln\phi)\bigg[u^cu^b\nabla_bS^a+u^au^b\nabla_bS^c
  -2g^{ac}\frac{\ud\ln m}{\ud\ln\phi}S^b\nabla_b\ln\phi\bigg].
\end{split}
\end{equation}
Again, the right hand side is evaluated along $\sigma_0(\tau)$.

In the Fermi normal coordinates, the spatial components of $A^\mu$ are given by
\begin{equation}\label{accobs}
  A^{\hat j}=-\frac{\ud\ln m}{\ud\ln \phi}\partial^{\hat j}\ln\phi,
\end{equation}
according to Eq.~(\ref{elnongeos}). By Eq.~\eqref{eq-fn-sjdd}, one obtains
\begin{equation}\label{cins}
  \frac{\ud^2S^{\hat j}}{\ud\tau^2}=-R_{\hat 0\hat k\hat 0}{}^{\hat j}S^{\hat k}-S^{\hat k}\partial_{\hat k}\left(\frac{\ud\ln m}{\ud\ln \phi}\partial^{\hat j}\ln\phi\right)
  +\left(\frac{\ud\ln m}{\ud\ln \phi}\right)^2(\partial^{\hat j}\ln\phi) S^{\hat k}\partial_{\hat k}\ln\phi.
\end{equation}
When the scalar field is not excited,i.e., $\phi=\phi_0$, a constant, Eq.~\eqref{cins} reduces to the geodesic deviation equation,
\begin{equation}\label{eq-usu-geodev}
  \frac{\ud^2S^{\hat j}}{\ud\tau^2}=-R_{\hat 0\hat k\hat 0}{}^{\hat j}S^{\hat k}.
\end{equation}
This is expected as vSEP is caused by a dynamical scalar field.
In the next section, Eq.~\eqref{cins} will be used to analyze the polarization content of GWs in Horndeski theory.

\section{The Polarizations of Gravitational Waves in Horndeski Gravity}\label{sec-pols}

In Ref.~\cite{Hou:2017bqj}, the GW solutions for Horndeski theory \cite{Horndeski:1974wa} in the vacuum background have been obtained.
The polarization content of the theory was also determined using the linearized geodesic deviation equation, as the vSEP was completely ignored.
In this section, the GW solution will be substituted into Eq.~(\ref{cins}) to take into account the effect of the scalar field on the trajectories of self-gravitating test particles.
This will lead to a different polarization content of GWs in Horndeski theory.

Now, one expands the fields around the flat background such that $g_{\mu\nu}=\eta_{\mu\nu}+h_{\mu\nu}$ and $\phi=\phi_0+\varphi$.
At the leading order, one obtains
\begin{equation}\label{flatre}
  G_{2(0,0)}=0,\quad G_{2(1,0)}=0.
\end{equation}
At the first order,  the linearized equations of motion can be written in the following form,
\begin{gather}
\label{eq-dcphgseq}
(\Box-m_s^2)\varphi = 0,\\
\label{eq-einhga}
\Box\tilde h_{\mu\nu} = 0,
\end{gather}
where the scalar field $\varphi$ is generally massive with the squared mass given by
\begin{equation}
  \label{msq}
m_s^2=-\frac{K_{(2,0)}}{K_{(0,1)}-2G_{3(1,0)}+3G_{4(1,0)}^2/G_{4(0,0)}},
\end{equation}
and $\tilde h_{\mu\nu}$ is an auxiliary field defined as
\begin{equation}
\label{auht}
\tilde h_{\mu\nu}=h_{\mu\nu}-\frac{1}{2}\eta_{\mu\nu}\eta^{\alpha\beta}h_{\alpha\beta}-\chi\eta_{\mu\nu}\varphi,
\end{equation}
with $\chi=\frac{G_{4(1,0)}}{G_{4(0,0)}}$.
Note that the transverse-traceless (TT) gauge $\partial_\mu \tilde h^{\mu\nu}=0$, $\eta^{\mu\nu}\tilde h_{\mu\nu}=0$ has been made.
A GW propagating in the $+z$ direction is given below
\begin{gather}
  \tilde h_{\mu\nu}=e_{\mu\nu}\cos\Omega(t-z), \label{wavesol-h}\\
  \varphi=\varphi_0\cos(\omega t-kz), \label{wavesol-s}
\end{gather}
where $\omega^2-k^2=m_s^2$ and the only nonvanishing components of tensor wave amplitude $e_{\mu\nu}$ are $e_{11}=-e_{22}$ and $e_{12}$. The coordinate system in which the TT gauge is chosen is called the TT coordinate system.

One is interested in studying the relative acceleration of two nearby particles which were at rest before the arrival of the GW.
Because of the presence of the GW induced by the scalar field, one expects $\sigma_0(\tau)$ to deviate from a straight line in the TT coordinates, so one assumes its 3-velocity is $\vec v$ and $u^\mu=u^0(1,\vec v)$.
The normalization of $u^a$ implies that
\begin{equation}\label{u0}
  u^0=1+\frac{1}{2}h_{00}+O(v^2).
\end{equation}
The acceleration of $\sigma_0(\tau)$ can be approximated as
\begin{equation}\label{accapp}
  A^\mu\approx-\frac{s}{\phi_0}(\eta^{\mu\nu}+\underline{u}^\mu\underline{u}^\nu)\nabla_\nu\varphi,
\end{equation}
with $s=(\ud\ln m/\ud\ln\phi)|_{\phi_0}$ called the sensitivity and $\underline u^\mu=(1,\vec 0)$ the background value. Written in component form, the acceleration is given by
\begin{gather}\label{accappcom}
  A^0=0,\\
  A^j=-\delta^j_3ks\frac{\varphi_0}{\phi_0}\sin(\omega t-kz).
\end{gather}
On the other hand, the left hand side of Eq.~(\ref{accapp}) is, in coordinate basis,
\begin{equation}\label{accappexp}
\begin{split}
  A^\mu=&\frac{\ud^2x^\mu}{\ud\tau^2}+\Gamma^\mu{}_{\rho\nu}\frac{\ud x^\rho}{\ud\tau}\frac{\ud x^\nu}{\ud\tau}\\
    \approx&(u^0)^2\left(\frac{\ud^2x^\mu}{\ud t^2}+\Gamma^\mu{}_{00}\right)+u^0\frac{\ud u^0}{\ud t}\frac{\ud x^\mu}{\ud t}.
\end{split}
\end{equation}
Consider a trivial motion, i.e., $x=y=0$.
Then one obtains
\begin{gather}
  v_3\approx-\frac{k}{2\omega}\left(\chi-\frac{2s}{\phi_0}\right)\varphi_0\cos\omega t,\label{eq-a3-sol-v}\\
  z\approx-\frac{k}{2\omega^2}\left(\chi-\frac{2s}{\phi_0}\right)\varphi_0\sin\omega t.\label{eq-a3-sol-z}
\end{gather}
Here, the initial position of $\sigma_0(\tau)$ is chosen to be $x_0=y_0=z_0=0$.
In addition,
\begin{equation}\label{eq-u0}
  u^0=\frac{\ud t}{\ud \tau}\approx1+\frac{1}{2}\chi\varphi_0\cos\omega t,
\end{equation}
according to Eq.~\eqref{u0}, which implies that
\begin{equation}\label{tauf}
  \tau\approx t-\frac{\chi\varphi_0}{2\omega}\sin\omega t.
\end{equation}
From this, one clearly sees that the TT coordinate system is not the proper reference frame for the observer $\sigma_0(\tau)$.

Therefore, the trajectory of $\sigma_0(\tau)$ in the TT coordinate system is described by
\begin{gather}
  \tau= t-\frac{\chi\varphi_0}{2\omega}\sin\omega t,\label{eq-sol-tau} \\
  x=y=0, \\
  z=-\frac{k}{2\omega^2}\left(\chi-\frac{2s}{\phi_0}\right)\varphi_0\sin\omega t,\label{eq-sol-z}
\end{gather}
up to the linear order.
Because of the scalar field, the observer oscillates with the same frequency of the GW in the TT coordinate system according to Eq.~\eqref{eq-sol-z}.
The time dilatation also oscillates by Eq.~\eqref{eq-sol-tau}.

In the limit of GR ($\chi=s=0$), the trajectory of $\sigma_0(\tau)$ is thus $t=\tau, x^j=0$ up to the linear order, i.e., a geodesic of the background metric. If vSEP is weak, i.e. $s\approx0$, the trajectory is
\begin{gather}\label{traj}
  \tau=t-\frac{\chi\varphi_0}{2\omega}\sin\omega t, \\
  x=y=0,\\
  z=-\frac{k\chi\varphi_0}{2\omega^2}\sin\omega t.
\end{gather}
This agrees with Ref.~\cite{Hou:2017bqj}.
Although the particle $\sigma_0(\tau)$ does not follow a geodesic of the background metric, it still travels along a geodesic of the full metric.

\subsection{The relative acceleration in the Fermi normal coordinates}

In this subsection, one obtains the relative acceleration in the Fermi normal coordinates using Eq.~(\ref{cins}).
This discussion will also reveal the polarization content of GWs.
The 4-velocity of the observer is
\begin{equation}\label{4vobs}
  u^a=(e_{\hat 0})^a=(1+h_{00}/2,0,0,v_3),
\end{equation}
so the following triad can be chosen,
\begin{gather}\label{triad}
  (e_{\hat 1})^a=(0,1-h_{11}/2,-h_{12}/2,0),\\
  (e_{\hat 2})^a=(0,-h_{12}/2,1-h_{22}/2,0), \\
  (e_{\hat 3})^a=(v_3,0,0,1+h_{00}/2).
\end{gather}
These basic vectors are Fermi-Walker transported and evaluated along $\sigma_0(\tau)$.
The dual basis is denoted as $\{(e^{\hat{\mu}})_a\}$ and $(e^{\hat \mu})_\nu\approx\delta^{\hat \mu}_\nu$ is sufficient.

Up to the linear order in perturbations, Eq.~\eqref{eq-fn-sjdd} is given by
\begin{equation}\label{eq-fn-sjdd-li}
  \frac{\ud^2 S^{\hat j}}{\ud t^2}=-R_{\hat 0\hat k\hat 0}{}^{\hat j}S^{\hat k}+S^{\hat k}\partial_{\hat k}A^{\hat j},
\end{equation}
since the acceleration $A^{\hat j}$ is of the linear order, and the last term in Eq.~\eqref{eq-fn-sjdd} should be dropped.
Normally, one has to find the Fermi normal coordinates explicitly \cite{1982NCimB..71...37F,1991NCimB.106..101F}.
However, the Fermi normal coordinates differ from the TT coordinates by  quantities of order one, and
the Riemann tensor and the 4-acceleration of the test particle are both of linear order,
so any changes in their components caused by the coordinate transformation are of the second order in perturbations.
Therefore, one only has to calculate the components of the Riemann tensor  and the 4-acceleration in the TT coordinates, and then simply substitutes them in Eq.~\eqref{eq-fn-sjdd-li}.

More explicitly, the driving force matrix is given by
\begin{equation}\label{drifm}
\begin{split}
  &S_{\hat k}{}^{\hat j}=R_{\hat 0\hat k\hat 0}{}^{\hat j}-\partial_{\hat k}A^{\hat j}\\
  &\approx R_{ 0 k 0}{}^{ j}-\partial_{ k}A^{ j} \\
 & \approx\left(
    \begin{array}{ccc}
      -\frac{\omega^2}{2}\chi\varphi+\frac{\Omega^2}{2}\tilde h_{11} & \frac{\Omega^2}{2}\tilde h_{12} & 0 \\
      \frac{\Omega^2}{2}\tilde h_{12} & -\frac{\omega^2}{2}\chi\varphi-\frac{\Omega^2}{2}\tilde h_{11} & 0 \\
      0 & 0 & -\frac{m_s^2}{2}\chi\varphi-\frac{k^2s}{\phi_0}\varphi
      \end{array}
      \right),
\end{split}
\end{equation}
where $\tilde h_{\mu\nu}$ and $\varphi$ are evaluated at $(t,\vec x=0)$. Comparing this matrix with the one (Eq.~(29)) in Ref.~\cite{Hou:2017bqj}, one finds out that vSEP introduces an order one correction $-k^2s\varphi/\phi_0$ to the longitudinal polarization.
This means that the longitudinal polarization gets enhanced.
Even if the scalar field is massless, the longitudinal polarization persists because the test particles are accelerated.

However, the enhancement is very extremely small for objects such as the mirrors used in detectors such as LIGO.
According to Refs.~\cite{Zaglauer1992,Alsing:2011er}, white dwarfs have typical sensitivities $s\sim10^{-4}$, so a test particle, like the mirror used by LIGO, would have an even smaller sensitivity.
So it would be still very difficult to use interferometers to detect the enhanced longitudinal polarization as in the previous case \cite{Hou:2017bqj}.
In contrast, neutron stars are compact objects.
Their sensitivity could be about 0.2 \cite{Zaglauer1992,Alsing:2011er}.
They violate SEP relatively strongly, which might be detected by PTAs.

\section{Pulsar Timing Arrays}\label{sec-pta}

In this section,  the cross-correlation function will be calculated for  PTAs.
The possibility to detect the vSEP is thus inferred.
%
A pulsar is a strongly magnetized, rotating neutron star or a white dwarf, which emits a beam of the radio wave along its magnetic pole.
When the beam points towards the Earth, the radiation is observed, and this leads to the pulsed appearance of the radiation.
The rotation of some ``recycled" pulsars is stable enough so that they can be used as ``cosmic light-house" \cite{Berti:2015itd}.
Among them, millisecond pulsars are found to be more stable \cite{2004hpa..book.....L} and used as stable clocks \cite{Verbiest:2009kb}.
When there is no GW, the radio pulses arrive at the Earth at a steady rate.
The presence of the GW will affect the propagation time of the radiation and thus alter this rate.
This results in a change in the time-of-arrival (TOA), called timing residual $R(t)$.
Timing residuals caused by the stochastic GW background is correlated between pulsars, and the cross-correlation function is $C(\theta)=\langle R_a(t)R_b(t)\rangle$ with $\theta$  the angular separation of pulsars $a$ and $b$,
and the brackets $\langle\,\rangle$ implying the ensemble average over the stochastic background.
This makes it possible to detect GWs and  probe the polarizations \cite{1975GReGr...6..439E,1978SvA....22...36S,Detweiler:1979wn,Hellings:1983fr,Jenet:2005pv,2008ApJ...685.1304L,Lee:2010cg,Lee:2014awa,Chamberlin:2011ev,Yunes:2013dva,Gair:2014rwa,Gair:2015hra}.
The effect of vSEP can also be detected, as the longitudinal polarization of the scalar-tensor theory is enhanced due to vSEP.

\begin{figure}[h]
\includegraphics[width=0.4\textwidth]{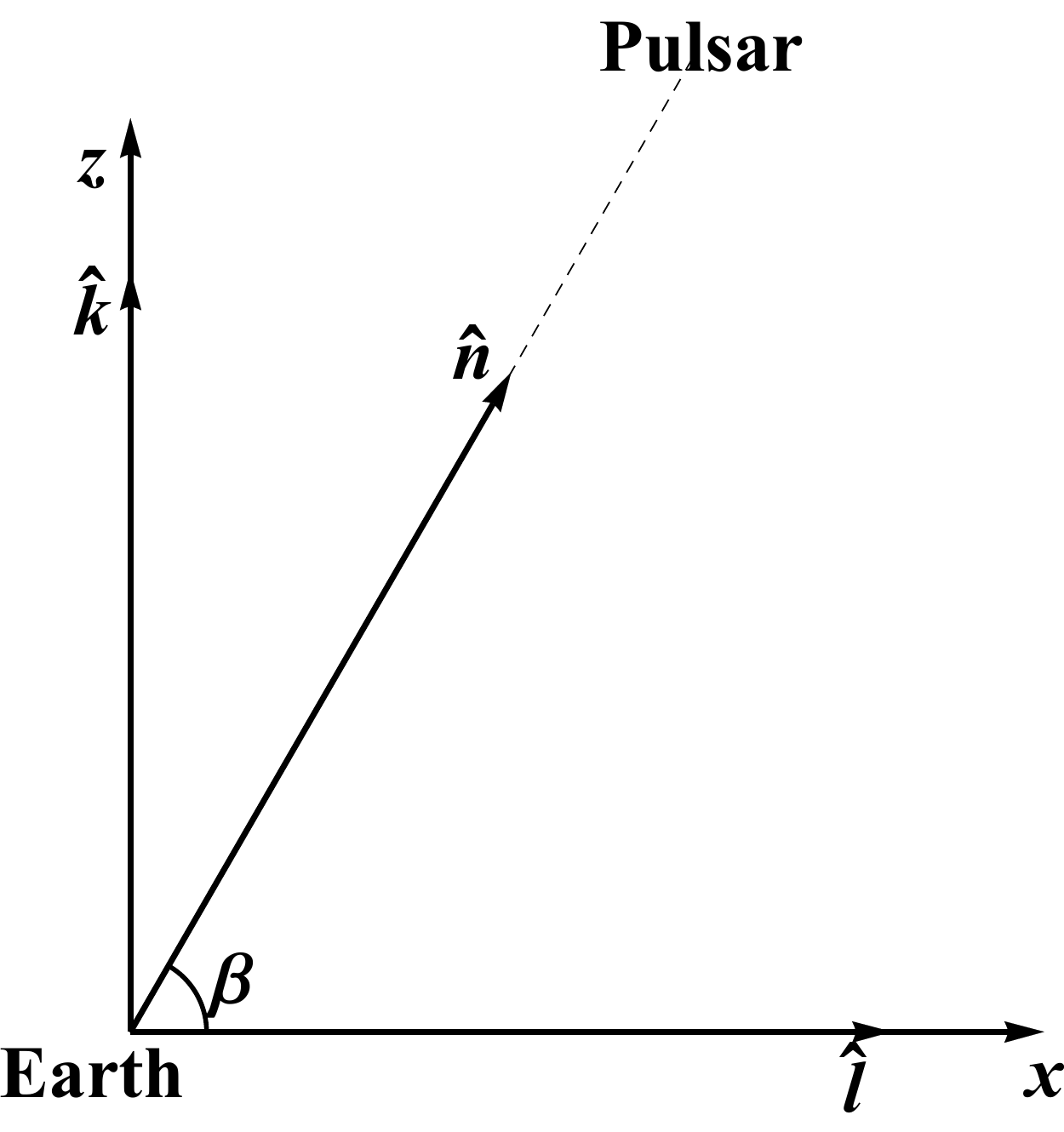}
\caption{The GW is propagating in the direction of $\hat k$, and the photon is traveling in $-\hat n$ direction at the leading order. $\hat l$ is perpendicular to $\hat k$ and in the same plane determined by $\hat k$ and $\hat n$. The angle between $\hat n$ and $\hat l$ is $\beta$.}\label{fig-coord}
\end{figure}

One sets up a coordinate system shown in Fig.~\ref{fig-coord} to calculate the timing residual $R(t)$ caused by the GW solution \eqref{wavesol-h} and \eqref{wavesol-s}.
Before the GW comes, the Earth is at the origin, and the distant pulsar is at rest at $\vec x_p=(L\cos\beta,0,L\sin\beta)$ in this coordinate system.
The GW is propagating in the direction of a unit vector $\hat k$, and $\hat n$ is the unit vector pointing  to the pulsar from the Earth.
$\hat l=\hat k\wedge(\hat n\wedge\hat k)/\cos\beta=[\hat n-\hat k(\hat{n}\cdot\hat k)]/\cos\beta$ is actually the unit vector parallel to the $y$ axis.
Where there is no GW, the photon is assumed to have a 4-velocity given by $\underline u^\mu=\gamma_0(1,-\cos\beta,0,-\sin\beta)$ with $\gamma_0=\ud t/\ud \lambda$ a constant and $\lambda$ an arbitrary affine parameter.
Let the perturbed photon 4-velocity be $u^\mu=\underline{u}^\mu+v^\mu$.
The condition $g_{\mu\nu}u^\mu u^\nu=0$ together with the photon geodesic equation lead to
\begin{eqnarray}
  v^0&=&\gamma_0\Big\{\chi\varphi_0\cos[(\omega+k\sin\beta)t-k(L+t_e)\sin\beta]\nonumber\\
  &&-\frac{e_{11}}{2}(1-\sin\beta)\cos[\Omega(1+\sin\beta)t-\Omega (L+t_e)\sin\beta]\Big\}, \label{eq-v0v}\\
  v_1&=&\gamma_0\{-\chi\varphi_0\cos\beta\cos[(\omega+k\sin\beta)t-k(L+t_e)\sin\beta]\nonumber\\
  &&+e_{11}\cos\beta\cos[\Omega(1+\sin\beta)t-\Omega (L+t_e)\sin\beta]\}, \label{eq-v1v}\\
  v_2&=&\gamma_0e_{12}\cos\beta\cos\Omega[(1+\sin\beta)t-(L+t_e)\sin\beta], \label{eq-v2v}\\
  v_3&=&\gamma_0\Big\{-\chi\varphi_0\sin\beta\cos[(\omega+k\sin\beta)t-k(L+t_e)\sin\beta]\nonumber\\
  &&-\frac{e_{11}}{2}(1-\sin\beta)\cos[\Omega(1+\sin\beta)t-\Omega (L+t_e)\sin\beta]\Big\},\label{eq-v3v}
\end{eqnarray}
where $t_e$ is the time when the photon is emitted from the pulsar.

The 4-velocity of an observer on the Earth has been obtained in Section~\ref{sec-pols}, which reads
\begin{equation}\label{eq-4vobs-e}
  T_e^\mu=\left(1+\frac{1}{2}\chi\varphi_0\cos\omega t,0,0,-\frac{k}{2\omega}\left(\chi-\frac{2s_r}{\phi_0}\right)\varphi_0\cos\omega t\right),
\end{equation}
where $s_r$ is the sensitivity of the Earth.
The 4-velocity of another observer comoving with the pulsar can be derived in a similar way. In fact, the translational symmetry in the background spacetime (i.e., Minkowskian spacetime) gives
\begin{equation}\label{eq-4vobs-p}
  T_p^\mu=\left(1+\frac{1}{2}\chi\varphi_0\cos(\omega t-kL\sin\beta),0,0,-\frac{k}{2\omega}\left(\chi-\frac{2s_e}{\phi_0}\right)\varphi_0\cos(\omega t-kL\sin\beta)\right),
\end{equation}
which agrees with the result from the direct calculation. Here, $s_e$ is the sensitivity of the pulsar.

So the measured frequency by the observer on the Earth is
\begin{equation}\label{eq-freqear}
  \begin{split}
  f_r=&-u_\mu T^\mu_e\\
  =&\gamma_0\Bigg[1+\left(\frac{\omega-k\sin\beta}{2\omega}\chi+\frac{s_rk}{\phi_0\omega}\sin\beta\right)\varphi_0
  \cos\omega(t_e+L)-\frac{e_{11}}{2}(1-\sin\beta)\cos\Omega(t_e+L)\Bigg],
  \end{split}
\end{equation}
and the one by the observer comoving with the pulsar is
\begin{equation}\label{eq-freqpul}
\begin{split}
  f_e=&-u_\mu T^\mu_p\\
  =&\gamma_0\Bigg[1+\left(\frac{\omega-k\sin\beta}{2\omega}\chi+\frac{s_ek}{\phi_0\omega}\sin\beta\right)\varphi_0\cos(\omega t_e
  -kL\sin\beta)\\
  &-\frac{e_{11}}{2}(1-\sin\beta)\cos\Omega(t_e-L\sin\beta)\Bigg].
  \end{split}
\end{equation}
Therefore, the frequency shift is given by
\begin{equation}\label{eq-fss}
  \begin{split}
  \frac{f_e-f_r}{f_r}=&\frac{\omega-k\hat k\cdot\hat n}{2\omega}\chi\left[\varphi(t-L,L\hat n)-\varphi(t,0)\right]\\
   &-\frac{e_{jk}\hat n^j\hat n^k}{2(1+\hat k\cdot\hat n)}\left[\tilde h_{jk}(t-L,L\hat n)-\tilde h_{jk}(t,0)\right]\\
    &+\frac{k}{\omega\phi_0}\hat k\cdot\hat n[s_e\varphi(t-L,L\hat n)-s_r\varphi(t,0)],
  \end{split}
\end{equation}
where $t=t_e+L$ is the time when the photon arrives at the Earth at the leading order.
This equation has been expressed in a coordinate independent way, so it can be straightforwardly used in any coordinate system with arbitrary orientation and at rest relative to the original one.
Note that the first two lines reproduce the result in Ref.~\cite{Hou:2017bqj}, and the third line comes from the effect of vSEP. This effect is completely determined by the scalar perturbation $\varphi$, as expected.

Therefore, the focus will be on the cross-correlation function for the scalar GW in the following discussion. Eq.~\eqref{eq-fss} is the frequency shift due to a monochromatic wave.
Now, consider the contribution of a stochastic GW background which consists of monochromatic GWs,
\begin{equation}
  \varphi(t,\vec x)=\int_{-\infty}^{\infty}\frac{\ud \omega}{2\pi}\int\ud^2\hat k \Big\{\varphi_0(\omega,\hat k )\exp[i(\omega t-k \hat k \cdot\vec x)]\Big\},
\end{equation}
where $\varphi_0(\omega ,\hat k)$ is the amplitude for the scalar GW propagating in the direction $\hat k $ at the angular frequency $\omega $.
Usually, one assumes that the GW background is isotropic, stationary and independently polarized, then one can define the characteristic strains $\varphi_c$ given by,
\begin{equation}
\langle \varphi_0^*(\omega ,\hat k )\varphi_0(\omega',\hat k')\rangle=\delta(\omega -\omega)\delta(\hat k -\hat k')\frac{|\varphi_c(\omega )|^2}{\omega },\label{defvc}
\end{equation}
where the star $*$ implies the complex conjugation.

The total timing residual in TOA due to the stochastic GW background is
\begin{equation}\label{rtsl}
  R(T)=\int_{-\infty}^{\infty}\frac{\ud \omega }{2\pi}\int\ud^2\hat k \int_{0}^{T}\ud t\frac{f_e-f_r}{f_r},
\end{equation}
where the argument $T$ is the total observation time. Insert Eq.~\eqref{eq-fss} in, neglecting the second line, to obtain
\begin{equation}\label{eq-rtslj}
\begin{split}
  R(T)=&\int_{-\infty}^{\infty}\frac{\ud \omega }{2\pi}\int\ud^2\hat k \varphi_0(\omega ,\hat k )(e^{i\omega T}-1)\left\{\frac{\omega -k \hat k \cdot\hat n}{i2\omega ^2}\chi\times\right.\\
  &\left.[e^{-i(\omega +k \hat k \cdot n)L}-1]+\frac{k\hat k\cdot\hat n}{i\omega^2\phi_0}[s_ee^{-i(\omega +k \hat k \cdot n)L}-s_r]\right\}.
  \end{split}
\end{equation}
With this result, consider the correlation between two pulsars $a$ and $b$  located at  $\vec x_a=L_1\hat n_1$ and $\vec x_b=L_2\hat n_2$, respectively. The angular separation is $\theta=\arccos(\hat n_1\cdot \hat n_2)$. The cross-correlation function is thus given by
\begin{equation}\label{eq-ccf}
\begin{split}
  C(\theta)=&\langle R_a(T)R_b(T)\rangle\\
  =&\int_{m_s}^\infty\ud\omega\int\ud^2\hat k\frac{|\varphi_c(\omega)|^2}{\pi\omega^5}\left[\frac{k^2\hat k\cdot\hat n_1\hat k\cdot\hat n_2}{\phi_0^2}\mathcal P_4+\frac{k\hat k\cdot\hat n_1(\omega-k\hat k\cdot\hat n_2)}{2\phi_0}\chi\mathcal P_2\right.\\
  &\left.+\frac{k\hat k\cdot\hat n_2(\omega-k\hat k\cdot\hat n_1)}{2\phi_0}\chi\mathcal P_3
  +\frac{(\omega-k\hat k\cdot\hat n_1)(\omega-k\hat{k}\cdot\hat{n}_2)}{4}\chi^2\mathcal{P}_1\right],
  \end{split}
\end{equation}
where $\mathcal P_1,\,\mathcal P_2,\,\mathcal P_3$ and $\mathcal P_4$ are defined to be
\begin{gather}
  \mathcal P_1=1-\cos\Delta_1-\cos\Delta_2+\cos(\Delta_1-\Delta_2),\label{eq-defps-1} \\
  \mathcal P_2=s_r-s_r\cos\Delta_2-s_e\cos\Delta_1+s_e\cos(\Delta_1-\Delta_2),\label{eq-defps-2} \\
  \mathcal P_3=s_r-s_r\cos\Delta_1-s_e\cos\Delta_2+s_e\cos(\Delta_1-\Delta_2),\label{eq-defps-3}\\
  \mathcal P_4=s_r^2-s_rs_e\cos\Delta_1-s_rs_e\cos\Delta_2+s_e^2\cos(\Delta_1-\Delta_2),\label{eq-defps-4}
\end{gather}
with $\Delta_j=(\omega+k \hat k \cdot \hat n_j)L_j$ for $j=1,2$.
To obtain this result,  Eq.~\eqref{defvc} is used, and the real part is taken. In addition, $T$ drops out, as the ensemble average also implies the averaging over the time \cite{2008ApJ...685.1304L}.

Because of the isotropy of the GW background, one sets
\begin{gather}\label{defns}
  \hat n_1=(0,0,1), \\
  \hat n_2=(\sin\theta,0,\cos\theta).
\end{gather}
Also, let $\hat k =(\sin\theta_g\cos\phi_g,\sin\theta_g\sin\phi_g,\cos\theta_g)$, so
\begin{gather}
  \Delta_1=(\omega +k \cos\theta_g)L_1,\\
  \Delta_2=[\omega +k (\sin\theta_g\cos\phi_g\sin\theta+\cos\theta_g\cos\theta)]L_2,
\end{gather}
Working in the limit that $\omega L_j\gg1$,  one can drop the cosines in  the definitions \eqref{eq-defps-1}-\eqref{eq-defps-4} of $\mathcal{P}_j$ ($j=1,2,3,4$), when $\theta\ne0$. The integration can be partially done, resulting in
\begin{equation}\label{eq-ccne0}
  C(\theta)=\int_{m_s}^\infty\ud\omega\frac{|\varphi_c(\omega)|^2}{\omega^3}\chi^2\left[1+\frac{k^2}{3\omega^2}\left(1-\frac{2s_r}{\phi_0\chi}\right)^2\cos\theta\right].
\end{equation}
But for $\theta=0$,  one considers the auto-correlation function, so set $\hat n_1=\hat n_2=(0,0,1)$ and $L_1=L_2=L$. The auto-correlation function is thus given by
\begin{equation}\label{eq-cceq0}
  C(0)=\int_{m_s}^\infty\ud\omega\frac{|\varphi_c(\omega)|^2}{\omega^3}\chi^2\left[2+\frac{k^2}{3\omega^2}\left(1-\frac{2s_r}{\phi_0\chi}\right)^2
  +\frac{k^2}{3\omega^2}\left(1-\frac{2s_e}{\phi_0\chi}\right)^2\right],
\end{equation}
where the terms containing $L$ are dropped as they barely contribute according to the experience in Ref.~\cite{Hou:2017bqj}.
Finally, the observation time $T$ sets a natural cutoff for the angular frequency, i.e., $\omega\ge 2\pi/T$, so the lower integration limits in Eqs.~\eqref{eq-ccne0} and \eqref{eq-cceq0} should be replaced by $\text{Max}\{m_s,2\pi/T\}$.

\begin{figure}
  \centering
  \includegraphics[width=0.48\textwidth]{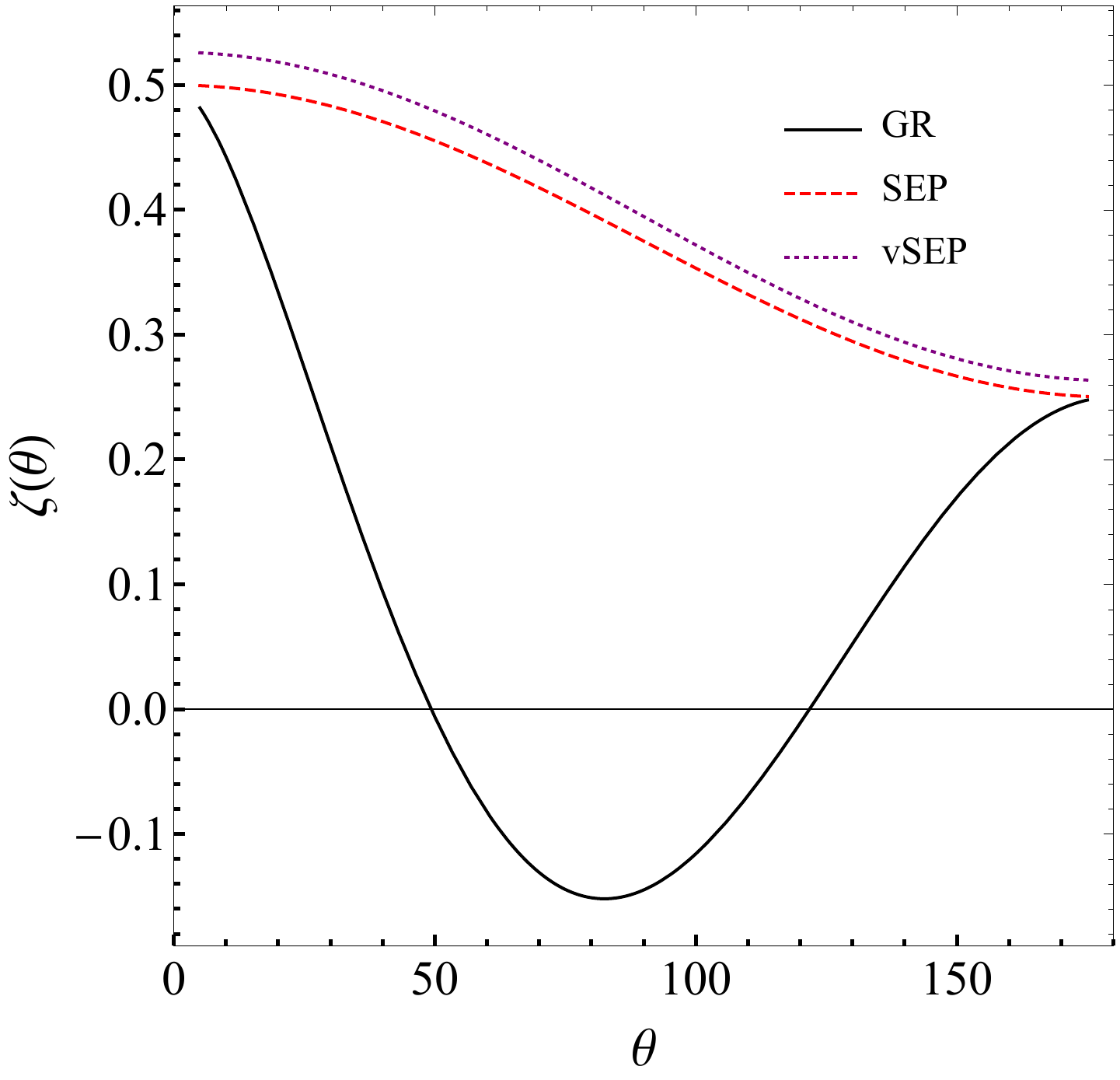}
  \includegraphics[width=0.48\textwidth]{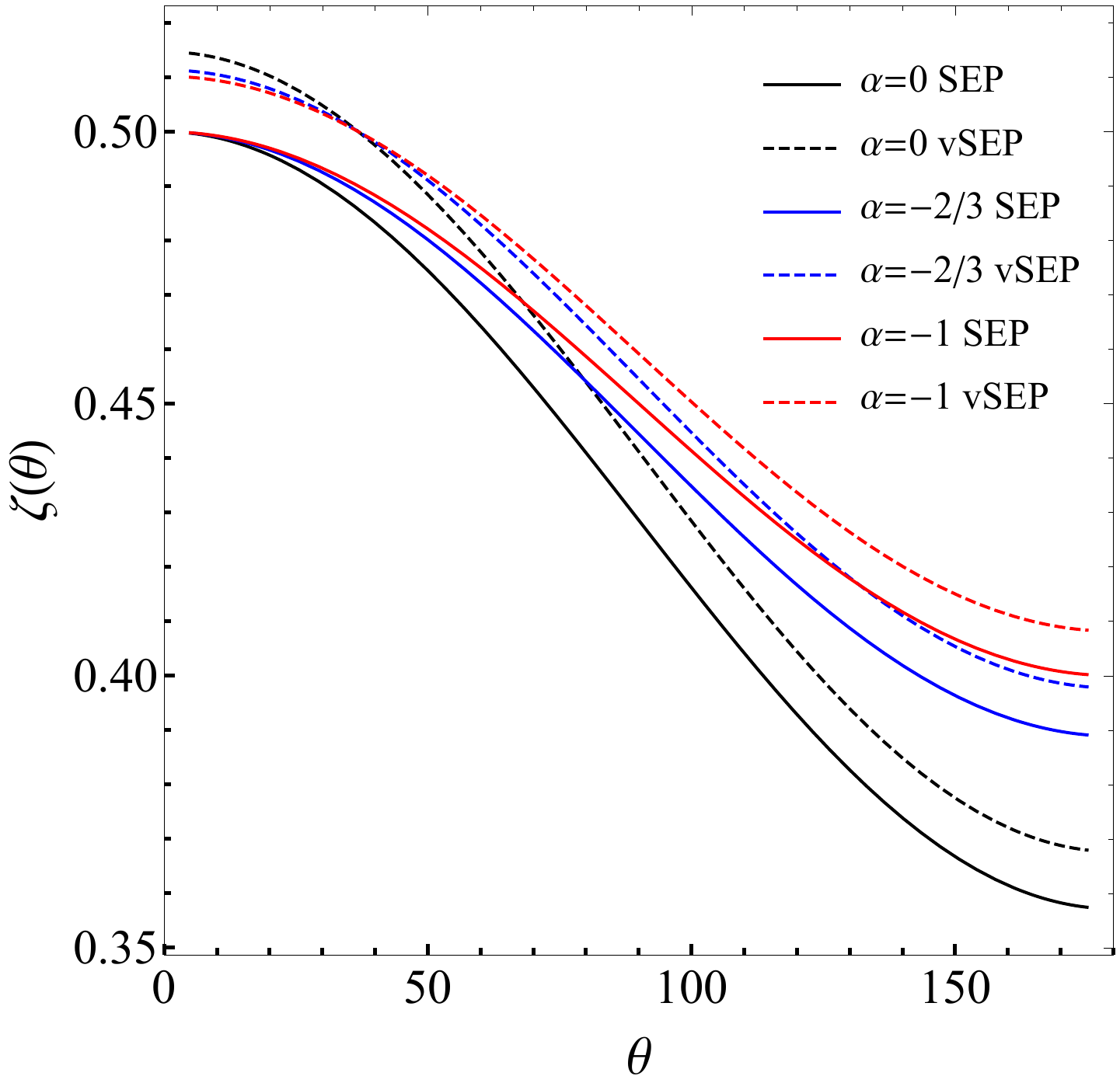}
  \caption{The normalized cross-correlation functions $\zeta(\theta)=C(\theta)/C(0)$.
  The left panel shows the cross-correlations when the scalar field is massless, i.e., when there is no longitudinal polarization.
  The solid curve is for familiar GR polarizations (i.e., the plus or cross ones), the dashed red curve for the breathing polarization with SEP and the dotted purple curve for the breathing polarization with vSEP.
  The right panel shows the normalized cross-correlations induced together by the transverse breathing and longitudinal polarizations when the mass of the scalar field is taken to be $m_s=7.7\times10^{-23}\,\mathrm{eV}/c^2$.
  The solid curves are for the cases where SEP is satisfied, while the dashed curves are for those where SEP is violated.
  The power-law index $\alpha=0,-2/3,-1$.
  The calculation was done assuming $T=5$ yrs.}\label{fig-grbrsen}
\end{figure}

As usual,  assume $\varphi_c(\omega)\propto(\omega /\omega_c)^\alpha$ with $\omega_c$ the characteristic angular frequency. Here, $\alpha$ is called the power-law index, and usually, $\alpha=0,\,-2/3$ or $-1$ \cite{2008ApJ...685.1304L,Romano:2016dpx}.
Numerically integrating Eqs.~\eqref{eq-ccne0} and \eqref{eq-cceq0} gives the so-called normalized correlation function $\zeta(\theta)=C(\theta)/C(0)$.
In the integration, set the observation time $T=5$ years.
The sensitivities of the Earth and the pulsar are taken to be $s_r=0$ and $s_e=0.2$, respectively.
This leads to  Fig.~\ref{fig-grbrsen}, where the power-law index $\alpha$ takes different values.

If the scalar field is massless, the results are shown in the left panel which displays  the normalized correlation functions for the plus and cross polarizations -- Hellings-Downs curve (labeled by ``GR") \cite{Hellings:1983fr}.
The remaining two curves are for the breathing polarization:
the dashed one is for the case where SEP is respected, while the dotted one is for the case where SEP is violated.
They are independent of the power-law index $\alpha$.
As one can see that vSEP makes $\zeta(\theta)$ bigger by about 5\%.
If the scalar field has a mass $m_s=7.7\times10^{-23}\,\mathrm{eV}/c^2$, the results are shown in the right panel.
In this panel, the cross-correlation functions for the scalar polarization are drawn for different values of $\alpha$.
The solid curves correspond to the case where SEP is satisfied, and the dashed curves are for the case where SEP is violated.
Since the cross correlation for the plus and  cross polarizations does not change, we do not plot them again in the right panel.
In the massive case, vSEP also increases $\zeta(\theta)$ by about 2\% to 3\%.

Ref.~\cite{Arzoumanian:2018saf} published the constraint on the stochastic GW background based on the recently released 11-year dataset from the North American Nanohertz Observatory for Gravitational Waves (NANOGrav).
Assuming the background is isotropic and $\alpha=-2/3$, the strain amplitude of the GW is less than $1.45\times10^{-15}$ at $f=1\text{ yr}^{-1}$.
In addition, the top panel in Figure 6 shows the observed cross correlation.
As one can clearly see, the error bars are very large\footnote{Precisely due to the large errors, we do not plot the observed cross correlation  in our Fig.~\ref{fig-grbrsen}.}.
More observations are needed to improve the statistics.

\section{Conclusion}\label{sec-con}

This work discusses the effects of the vSEP on the polarization content of GWs in Horndeski theory and calculates the cross-correlation functions for PTAs.
Because of the vSEP, self-gravitating particles no longer travel along geodesics, and this leads to the enhancement of the longitudinal polarization in Horndeski theory,
so even if the scalar field is massless, the longitudinal polarization still exists.
This is in contrast with the previous results \cite{Hou:2017bqj,Liang:2017ahj,Gong:2017bru,Gong:2018ybk} that the massive scalar field excites the longitudinal polarization, while the massless scalar field does not.
The enhanced longitudinal polarization is nevertheless difficult for aLIGO to detect, as the mirrors does not violate SEP enough.
However, pulsars are highly compact objects with sufficient self-gravitating energy such that their trajectories deviate from geodesics enough.
Using PTAs, one can measure the change in TOAs of electromagnetic radiation from pulsars and obtain the cross-correlation function to tell whether vSEP effect exits.
The results show that the vSEP leads to large changes in the behaviors of the cross-correlation functions.
In principle, PTAs are capable of detecting the vSEP if it exists.

\begin{acknowledgements}
This research was supported in part by the Major Program of the National Natural Science Foundation of China under Grant No. 11475065 and the National Natural Science Foundation of China under Grant No. 11690021.
This was also a project funded by China Postdoctoral Science Foundation (No. 2018M632822).
\end{acknowledgements}


\begin{thebibliography}{80}%
\makeatletter
\providecommand \@ifxundefined [1]{%
 \@ifx{#1\undefined}
}%
\providecommand \@ifnum [1]{%
 \ifnum #1\expandafter \@firstoftwo
 \else \expandafter \@secondoftwo
 \fi
}%
\providecommand \@ifx [1]{%
 \ifx #1\expandafter \@firstoftwo
 \else \expandafter \@secondoftwo
 \fi
}%
\providecommand \natexlab [1]{#1}%
\providecommand \enquote  [1]{``#1''}%
\providecommand \bibnamefont  [1]{#1}%
\providecommand \bibfnamefont [1]{#1}%
\providecommand \citenamefont [1]{#1}%
\providecommand \href@noop [0]{\@secondoftwo}%
\providecommand \href [0]{\begingroup \@sanitize@url \@href}%
\providecommand \@href[1]{\@@startlink{#1}\@@href}%
\providecommand \@@href[1]{\endgroup#1\@@endlink}%
\providecommand \@sanitize@url [0]{\catcode `\\12\catcode `\$12\catcode
  `\&12\catcode `\#12\catcode `\^12\catcode `\_12\catcode `\%12\relax}%
\providecommand \@@startlink[1]{}%
\providecommand \@@endlink[0]{}%
\providecommand \url  [0]{\begingroup\@sanitize@url \@url }%
\providecommand \@url [1]{\endgroup\@href {#1}{\urlprefix }}%
\providecommand \urlprefix  [0]{URL }%
\providecommand \Eprint [0]{\href }%
\providecommand \doibase [0]{http://dx.doi.org/}%
\providecommand \selectlanguage [0]{\@gobble}%
\providecommand \bibinfo  [0]{\@secondoftwo}%
\providecommand \bibfield  [0]{\@secondoftwo}%
\providecommand \translation [1]{[#1]}%
\providecommand \BibitemOpen [0]{}%
\providecommand \bibitemStop [0]{}%
\providecommand \bibitemNoStop [0]{.\EOS\space}%
\providecommand \EOS [0]{\spacefactor3000\relax}%
\providecommand \BibitemShut  [1]{\csname bibitem#1\endcsname}%
\let\auto@bib@innerbib\@empty
\bibitem [{\citenamefont {Perlmutter}\ \emph {et~al.}(1999)\citenamefont
  {Perlmutter} \emph {et~al.}}]{Perlmutter:1998np}%
  \BibitemOpen
  \bibfield  {author} {\bibinfo {author} {\bibfnamefont {S.}~\bibnamefont
  {Perlmutter}} \emph {et~al.} (\bibinfo {collaboration} {Supernova Cosmology
  Project}),\ }\href {\doibase 10.1086/307221} {\bibfield  {journal} {\bibinfo
  {journal} {Astrophys. J.}\ }\textbf {\bibinfo {volume} {517}},\ \bibinfo
  {pages} {565} (\bibinfo {year} {1999})},\ \Eprint
  {http://arxiv.org/abs/astro-ph/9812133} {arXiv:astro-ph/9812133 [astro-ph]}
  \BibitemShut {NoStop}%
\bibitem [{\citenamefont {Riess}\ \emph {et~al.}(1998)\citenamefont {Riess}
  \emph {et~al.}}]{Riess:1998cb}%
  \BibitemOpen
  \bibfield  {author} {\bibinfo {author} {\bibfnamefont {A.~G.}\ \bibnamefont
  {Riess}} \emph {et~al.} (\bibinfo {collaboration} {Supernova Search Team}),\
  }\href {\doibase 10.1086/300499} {\bibfield  {journal} {\bibinfo  {journal}
  {Astron. J.}\ }\textbf {\bibinfo {volume} {116}},\ \bibinfo {pages} {1009}
  (\bibinfo {year} {1998})},\ \Eprint {http://arxiv.org/abs/astro-ph/9805201}
  {arXiv:astro-ph/9805201 [astro-ph]} \BibitemShut {NoStop}%
\bibitem [{\citenamefont {Abbott}\ \emph
  {et~al.}(2016{\natexlab{a}})\citenamefont {Abbott} \emph
  {et~al.}}]{Abbott:2016blz}%
  \BibitemOpen
  \bibfield  {author} {\bibinfo {author} {\bibfnamefont {B.~P.}\ \bibnamefont
  {Abbott}} \emph {et~al.} (\bibinfo {collaboration} {Virgo, LIGO
  Scientific}),\ }\href {\doibase 10.1103/PhysRevLett.116.061102} {\bibfield
  {journal} {\bibinfo  {journal} {Phys. Rev. Lett.}\ }\textbf {\bibinfo
  {volume} {116}},\ \bibinfo {pages} {061102} (\bibinfo {year}
  {2016}{\natexlab{a}})},\ \Eprint {http://arxiv.org/abs/1602.03837}
  {arXiv:1602.03837 [gr-qc]} \BibitemShut {NoStop}%
\bibitem [{\citenamefont {Abbott}\ \emph
  {et~al.}(2016{\natexlab{b}})\citenamefont {Abbott} \emph
  {et~al.}}]{Abbott:2016nmj}%
  \BibitemOpen
  \bibfield  {author} {\bibinfo {author} {\bibfnamefont {B.~P.}\ \bibnamefont
  {Abbott}} \emph {et~al.} (\bibinfo {collaboration} {Virgo, LIGO
  Scientific}),\ }\href {\doibase 10.1103/PhysRevLett.116.241103} {\bibfield
  {journal} {\bibinfo  {journal} {Phys. Rev. Lett.}\ }\textbf {\bibinfo
  {volume} {116}},\ \bibinfo {pages} {241103} (\bibinfo {year}
  {2016}{\natexlab{b}})},\ \Eprint {http://arxiv.org/abs/1606.04855}
  {arXiv:1606.04855 [gr-qc]} \BibitemShut {NoStop}%
\bibitem [{\citenamefont {Abbott}\ \emph
  {et~al.}(2017{\natexlab{a}})\citenamefont {Abbott} \emph
  {et~al.}}]{Abbott:2017vtc}%
  \BibitemOpen
  \bibfield  {author} {\bibinfo {author} {\bibfnamefont {B.~P.}\ \bibnamefont
  {Abbott}} \emph {et~al.} (\bibinfo {collaboration} {Virgo, LIGO
  Scientific}),\ }\href {\doibase 10.1103/PhysRevLett.118.221101} {\bibfield
  {journal} {\bibinfo  {journal} {Phys. Rev. Lett.}\ }\textbf {\bibinfo
  {volume} {118}},\ \bibinfo {pages} {221101} (\bibinfo {year}
  {2017}{\natexlab{a}})},\ \Eprint {http://arxiv.org/abs/1706.01812}
  {arXiv:1706.01812 [gr-qc]} \BibitemShut {NoStop}%
\bibitem [{\citenamefont {Abbott}\ \emph
  {et~al.}(2017{\natexlab{b}})\citenamefont {Abbott} \emph
  {et~al.}}]{Abbott:2017oio}%
  \BibitemOpen
  \bibfield  {author} {\bibinfo {author} {\bibfnamefont {B.~P.}\ \bibnamefont
  {Abbott}} \emph {et~al.} (\bibinfo {collaboration} {Virgo, LIGO
  Scientific}),\ }\href {\doibase 10.1103/PhysRevLett.119.141101} {\bibfield
  {journal} {\bibinfo  {journal} {Phys. Rev. Lett.}\ }\textbf {\bibinfo
  {volume} {119}},\ \bibinfo {pages} {141101} (\bibinfo {year}
  {2017}{\natexlab{b}})},\ \Eprint {http://arxiv.org/abs/1709.09660}
  {arXiv:1709.09660 [gr-qc]} \BibitemShut {NoStop}%
\bibitem [{\citenamefont {Abbott}\ \emph
  {et~al.}(2017{\natexlab{c}})\citenamefont {Abbott} \emph
  {et~al.}}]{TheLIGOScientific:2017qsa}%
  \BibitemOpen
  \bibfield  {author} {\bibinfo {author} {\bibfnamefont {B.~P.}\ \bibnamefont
  {Abbott}} \emph {et~al.} (\bibinfo {collaboration} {Virgo, LIGO
  Scientific}),\ }\href {\doibase 10.1103/PhysRevLett.119.161101} {\bibfield
  {journal} {\bibinfo  {journal} {Phys. Rev. Lett.}\ }\textbf {\bibinfo
  {volume} {119}},\ \bibinfo {pages} {161101} (\bibinfo {year}
  {2017}{\natexlab{c}})},\ \Eprint {http://arxiv.org/abs/1710.05832}
  {arXiv:1710.05832 [gr-qc]} \BibitemShut {NoStop}%
\bibitem [{\citenamefont {Abbott}\ \emph
  {et~al.}(2017{\natexlab{d}})\citenamefont {Abbott} \emph
  {et~al.}}]{Abbott:2017gyy}%
  \BibitemOpen
  \bibfield  {author} {\bibinfo {author} {\bibfnamefont {B.~P.}\ \bibnamefont
  {Abbott}} \emph {et~al.} (\bibinfo {collaboration} {Virgo, LIGO
  Scientific}),\ }\href {\doibase 10.3847/2041-8213/aa9f0c} {\bibfield
  {journal} {\bibinfo  {journal} {Astrophys. J.}\ }\textbf {\bibinfo {volume}
  {851}},\ \bibinfo {pages} {L35} (\bibinfo {year} {2017}{\natexlab{d}})},\
  \Eprint {http://arxiv.org/abs/1711.05578} {arXiv:1711.05578 [astro-ph.HE]}
  \BibitemShut {NoStop}%
\bibitem [{\citenamefont {Abbott}\ \emph
  {et~al.}(2018{\natexlab{a}})\citenamefont {Abbott} \emph
  {et~al.}}]{LIGOScientific:2018mvr}%
  \BibitemOpen
  \bibfield  {author} {\bibinfo {author} {\bibfnamefont {B.~P.}\ \bibnamefont
  {Abbott}} \emph {et~al.} (\bibinfo {collaboration} {LIGO Scientific,
  Virgo}),\ }\href@noop {} {\  (\bibinfo {year} {2018}{\natexlab{a}})},\
  \Eprint {http://arxiv.org/abs/1811.12907} {arXiv:1811.12907 [astro-ph.HE]}
  \BibitemShut {NoStop}%
\bibitem [{\citenamefont {Abbott}\ \emph
  {et~al.}(2018{\natexlab{b}})\citenamefont {Abbott} \emph
  {et~al.}}]{Abbott:2018lct}%
  \BibitemOpen
  \bibfield  {author} {\bibinfo {author} {\bibfnamefont {B.~P.}\ \bibnamefont
  {Abbott}} \emph {et~al.} (\bibinfo {collaboration} {LIGO Scientific,
  Virgo}),\ }\href@noop {} {\  (\bibinfo {year} {2018}{\natexlab{b}})},\
  \Eprint {http://arxiv.org/abs/1811.00364} {arXiv:1811.00364 [gr-qc]}
  \BibitemShut {NoStop}%
\bibitem [{\citenamefont {{Kramer}}\ and\ \citenamefont
  {{Champion}}(2013)}]{Kramer:2013kea}%
  \BibitemOpen
  \bibfield  {author} {\bibinfo {author} {\bibfnamefont {M.}~\bibnamefont
  {{Kramer}}}\ and\ \bibinfo {author} {\bibfnamefont {D.~J.}\ \bibnamefont
  {{Champion}}},\ }\href {\doibase 10.1088/0264-9381/30/22/224009} {\bibfield
  {journal} {\bibinfo  {journal} {Class. Quant. Grav.}\ }\textbf {\bibinfo
  {volume} {30}},\ \bibinfo {eid} {224009} (\bibinfo {year}
  {2013})}\BibitemShut {NoStop}%
\bibitem [{\citenamefont {Hobbs}\ \emph {et~al.}(2010)\citenamefont {Hobbs}
  \emph {et~al.}}]{Hobbs:2009yy}%
  \BibitemOpen
  \bibfield  {author} {\bibinfo {author} {\bibfnamefont {G.}~\bibnamefont
  {Hobbs}} \emph {et~al.},\ }\bibfield  {booktitle} {\emph {\bibinfo
  {booktitle} {{Gravitational waves. Proceedings, 8th Edoardo Amaldi
  Conference, Amaldi 8, New York, USA, June 22-26, 2009}}},\ }\href {\doibase
  10.1088/0264-9381/27/8/084013} {\bibfield  {journal} {\bibinfo  {journal}
  {Class. Quant. Grav.}\ }\textbf {\bibinfo {volume} {27}},\ \bibinfo {pages}
  {084013} (\bibinfo {year} {2010})},\ \Eprint {http://arxiv.org/abs/0911.5206}
  {arXiv:0911.5206 [astro-ph.SR]} \BibitemShut {NoStop}%
\bibitem [{\citenamefont {McLaughlin}(2013)}]{McLaughlin:2013ira}%
  \BibitemOpen
  \bibfield  {author} {\bibinfo {author} {\bibfnamefont {M.~A.}\ \bibnamefont
  {McLaughlin}},\ }\href {\doibase 10.1088/0264-9381/30/22/224008} {\bibfield
  {journal} {\bibinfo  {journal} {Class. Quant. Grav.}\ }\textbf {\bibinfo
  {volume} {30}},\ \bibinfo {pages} {224008} (\bibinfo {year} {2013})},\
  \Eprint {http://arxiv.org/abs/1310.0758} {arXiv:1310.0758 [astro-ph.IM]}
  \BibitemShut {NoStop}%
\bibitem [{\citenamefont {Hobbs}(2013)}]{Hobbs:2013aka}%
  \BibitemOpen
  \bibfield  {author} {\bibinfo {author} {\bibfnamefont {G.}~\bibnamefont
  {Hobbs}},\ }\href {\doibase 10.1088/0264-9381/30/22/224007} {\bibfield
  {journal} {\bibinfo  {journal} {Class. Quant. Grav.}\ }\textbf {\bibinfo
  {volume} {30}},\ \bibinfo {pages} {224007} (\bibinfo {year} {2013})},\
  \Eprint {http://arxiv.org/abs/1307.2629} {arXiv:1307.2629 [astro-ph.IM]}
  \BibitemShut {NoStop}%
\bibitem [{\citenamefont {Will}(1993)}]{Will:1993ns}%
  \BibitemOpen
  \bibfield  {author} {\bibinfo {author} {\bibfnamefont {C.~M.}\ \bibnamefont
  {Will}},\ }\href@noop {} {\emph {\bibinfo {title} {{Theory and experiment in
  gravitational physics}}}}\ (\bibinfo  {publisher} {Cambridge, UK: Univ. Pr.
  (1993) 380 p},\ \bibinfo {year} {1993})\BibitemShut {NoStop}%
\bibitem [{Note1()}]{Note1}%
  \BibitemOpen
  \bibinfo {note} {Nordstrom's scalar theory of gravity satisfies SEP, but is
  not a metric theory, not to mention that it has been excluded by the
  observations \cite {Deruelle:2011wu}.}\BibitemShut {Stop}%
\bibitem [{\citenamefont {Barausse}\ and\ \citenamefont
  {Yagi}(2015)}]{Barausse:2015wia}%
  \BibitemOpen
  \bibfield  {author} {\bibinfo {author} {\bibfnamefont {E.}~\bibnamefont
  {Barausse}}\ and\ \bibinfo {author} {\bibfnamefont {K.}~\bibnamefont
  {Yagi}},\ }\href {\doibase 10.1103/PhysRevLett.115.211105} {\bibfield
  {journal} {\bibinfo  {journal} {Phys. Rev. Lett.}\ }\textbf {\bibinfo
  {volume} {115}},\ \bibinfo {pages} {211105} (\bibinfo {year} {2015})},\
  \Eprint {http://arxiv.org/abs/1509.04539} {arXiv:1509.04539 [gr-qc]}
  \BibitemShut {NoStop}%
\bibitem [{\citenamefont {Brans}\ and\ \citenamefont
  {Dicke}(1961)}]{Brans:1961sx}%
  \BibitemOpen
  \bibfield  {author} {\bibinfo {author} {\bibfnamefont {C.}~\bibnamefont
  {Brans}}\ and\ \bibinfo {author} {\bibfnamefont {R.~H.}\ \bibnamefont
  {Dicke}},\ }\href {https://link.aps.org/doi/10.1103/PhysRev.124.925}
  {\bibfield  {journal} {\bibinfo  {journal} {Phys. Rev.}\ }\textbf {\bibinfo
  {volume} {124}},\ \bibinfo {pages} {925} (\bibinfo {year}
  {1961})}\BibitemShut {NoStop}%
\bibitem [{\citenamefont {Kanti}\ \emph {et~al.}(1996)\citenamefont {Kanti},
  \citenamefont {Mavromatos}, \citenamefont {Rizos}, \citenamefont {Tamvakis},\
  and\ \citenamefont {Winstanley}}]{Kanti:1995vq}%
  \BibitemOpen
  \bibfield  {author} {\bibinfo {author} {\bibfnamefont {P.}~\bibnamefont
  {Kanti}}, \bibinfo {author} {\bibfnamefont {N.~E.}\ \bibnamefont
  {Mavromatos}}, \bibinfo {author} {\bibfnamefont {J.}~\bibnamefont {Rizos}},
  \bibinfo {author} {\bibfnamefont {K.}~\bibnamefont {Tamvakis}}, \ and\
  \bibinfo {author} {\bibfnamefont {E.}~\bibnamefont {Winstanley}},\ }\href
  {\doibase 10.1103/PhysRevD.54.5049} {\bibfield  {journal} {\bibinfo
  {journal} {Phys. Rev. D}\ }\textbf {\bibinfo {volume} {54}},\ \bibinfo
  {pages} {5049} (\bibinfo {year} {1996})},\ \Eprint
  {http://arxiv.org/abs/hep-th/9511071} {arXiv:hep-th/9511071 [hep-th]}
  \BibitemShut {NoStop}%
\bibitem [{\citenamefont {Buchdahl}(1970)}]{Buchdahl:1983zz}%
  \BibitemOpen
  \bibfield  {author} {\bibinfo {author} {\bibfnamefont {H.~A.}\ \bibnamefont
  {Buchdahl}},\ }\href@noop {} {\bibfield  {journal} {\bibinfo  {journal} {Mon.
  Not. Roy. Astron. Soc.}\ }\textbf {\bibinfo {volume} {150}},\ \bibinfo
  {pages} {1} (\bibinfo {year} {1970})}\BibitemShut {NoStop}%
\bibitem [{\citenamefont {O'Hanlon}(1972)}]{OHanlon:1972xqa}%
  \BibitemOpen
  \bibfield  {author} {\bibinfo {author} {\bibfnamefont {J.}~\bibnamefont
  {O'Hanlon}},\ }\href {https://link.aps.org/doi/10.1103/PhysRevLett.29.137}
  {\bibfield  {journal} {\bibinfo  {journal} {Phys. Rev. Lett.}\ }\textbf
  {\bibinfo {volume} {29}},\ \bibinfo {pages} {137} (\bibinfo {year}
  {1972})}\BibitemShut {NoStop}%
\bibitem [{\citenamefont {Teyssandier}\ and\ \citenamefont
  {Tourrenc}(1983)}]{Teyssandier:1983zz}%
  \BibitemOpen
  \bibfield  {author} {\bibinfo {author} {\bibfnamefont {P.}~\bibnamefont
  {Teyssandier}}\ and\ \bibinfo {author} {\bibfnamefont {P.}~\bibnamefont
  {Tourrenc}},\ }\href {\doibase 10.1063/1.525659} {\bibfield  {journal}
  {\bibinfo  {journal} {J. Math. Phys.}\ }\textbf {\bibinfo {volume} {24}},\
  \bibinfo {pages} {2793} (\bibinfo {year} {1983})}\BibitemShut {NoStop}%
\bibitem [{\citenamefont {Horndeski}(1974)}]{Horndeski:1974wa}%
  \BibitemOpen
  \bibfield  {author} {\bibinfo {author} {\bibfnamefont {G.~W.}\ \bibnamefont
  {Horndeski}},\ }\href {\doibase 10.1007/BF01807638} {\bibfield  {journal}
  {\bibinfo  {journal} {Int. J. Theor. Phys.}\ }\textbf {\bibinfo {volume}
  {10}},\ \bibinfo {pages} {363} (\bibinfo {year} {1974})}\BibitemShut
  {NoStop}%
\bibitem [{\citenamefont {Ostrogradsky}(1850)}]{Ostrogradsky:1850fid}%
  \BibitemOpen
  \bibfield  {author} {\bibinfo {author} {\bibfnamefont {M.}~\bibnamefont
  {Ostrogradsky}},\ }\href@noop {} {\bibfield  {journal} {\bibinfo  {journal}
  {Mem. Acad. St. Petersbourg}\ }\textbf {\bibinfo {volume} {6}},\ \bibinfo
  {pages} {385} (\bibinfo {year} {1850})}\BibitemShut {NoStop}%
\bibitem [{\citenamefont {Nordtvedt}(1968{\natexlab{a}})}]{Nordtvedt:1968qr}%
  \BibitemOpen
  \bibfield  {author} {\bibinfo {author} {\bibfnamefont {K.}~\bibnamefont
  {Nordtvedt}},\ }\href {\doibase 10.1103/PhysRev.169.1014} {\bibfield
  {journal} {\bibinfo  {journal} {Phys. Rev.}\ }\textbf {\bibinfo {volume}
  {169}},\ \bibinfo {pages} {1014} (\bibinfo {year}
  {1968}{\natexlab{a}})}\BibitemShut {NoStop}%
\bibitem [{\citenamefont {Nordtvedt}(1968{\natexlab{b}})}]{Nordtvedt:1968qs}%
  \BibitemOpen
  \bibfield  {author} {\bibinfo {author} {\bibfnamefont {K.}~\bibnamefont
  {Nordtvedt}},\ }\href {\doibase 10.1103/PhysRev.169.1017} {\bibfield
  {journal} {\bibinfo  {journal} {Phys. Rev.}\ }\textbf {\bibinfo {volume}
  {169}},\ \bibinfo {pages} {1017} (\bibinfo {year}
  {1968}{\natexlab{b}})}\BibitemShut {NoStop}%
\bibitem [{\citenamefont {Nordtvedt}(1982)}]{1982RPPh...45..631N}%
  \BibitemOpen
  \bibfield  {author} {\bibinfo {author} {\bibfnamefont {K.}~\bibnamefont
  {Nordtvedt}},\ }\href {http://stacks.iop.org/0034-4885/45/i=6/a=002}
  {\bibfield  {journal} {\bibinfo  {journal} {Rep. Prog. Phys.}\ }\textbf
  {\bibinfo {volume} {45}},\ \bibinfo {pages} {631} (\bibinfo {year}
  {1982})}\BibitemShut {NoStop}%
\bibitem [{\citenamefont {Will}(2014)}]{Will:2014kxa}%
  \BibitemOpen
  \bibfield  {author} {\bibinfo {author} {\bibfnamefont {C.~M.}\ \bibnamefont
  {Will}},\ }\href {\doibase 10.12942/lrr-2014-4} {\bibfield  {journal}
  {\bibinfo  {journal} {Living Rev. Rel.}\ }\textbf {\bibinfo {volume} {17}},\
  \bibinfo {pages} {4} (\bibinfo {year} {2014})},\ \Eprint
  {http://arxiv.org/abs/1403.7377} {arXiv:1403.7377 [gr-qc]} \BibitemShut
  {NoStop}%
\bibitem [{\citenamefont {Alsing}\ \emph {et~al.}(2012)\citenamefont {Alsing},
  \citenamefont {Berti}, \citenamefont {Will},\ and\ \citenamefont
  {Zaglauer}}]{Alsing:2011er}%
  \BibitemOpen
  \bibfield  {author} {\bibinfo {author} {\bibfnamefont {J.}~\bibnamefont
  {Alsing}}, \bibinfo {author} {\bibfnamefont {E.}~\bibnamefont {Berti}},
  \bibinfo {author} {\bibfnamefont {C.~M.}\ \bibnamefont {Will}}, \ and\
  \bibinfo {author} {\bibfnamefont {H.}~\bibnamefont {Zaglauer}},\ }\href
  {\doibase 10.1103/PhysRevD.85.064041} {\bibfield  {journal} {\bibinfo
  {journal} {Phys. Rev. D}\ }\textbf {\bibinfo {volume} {85}},\ \bibinfo
  {pages} {064041} (\bibinfo {year} {2012})},\ \Eprint
  {http://arxiv.org/abs/1112.4903} {arXiv:1112.4903 [gr-qc]} \BibitemShut
  {NoStop}%
\bibitem [{\citenamefont {{Hofmann}}\ \emph {et~al.}(2010)\citenamefont
  {{Hofmann}}, \citenamefont {{M{\"u}ller}},\ and\ \citenamefont
  {{Biskupek}}}]{2010A&A...522L...5H}%
  \BibitemOpen
  \bibfield  {author} {\bibinfo {author} {\bibfnamefont {F.}~\bibnamefont
  {{Hofmann}}}, \bibinfo {author} {\bibfnamefont {J.}~\bibnamefont
  {{M{\"u}ller}}}, \ and\ \bibinfo {author} {\bibfnamefont {L.}~\bibnamefont
  {{Biskupek}}},\ }\href {\doibase 10.1051/0004-6361/201015659} {\bibfield
  {journal} {\bibinfo  {journal} {Astron. Astrophys.}\ }\textbf {\bibinfo
  {volume} {522}},\ \bibinfo {eid} {L5} (\bibinfo {year} {2010})}\BibitemShut
  {NoStop}%
\bibitem [{\citenamefont {Damour}\ and\ \citenamefont
  {Sch\"aefer}(1991)}]{Damour:1991rq}%
  \BibitemOpen
  \bibfield  {author} {\bibinfo {author} {\bibfnamefont {T.}~\bibnamefont
  {Damour}}\ and\ \bibinfo {author} {\bibfnamefont {G.}~\bibnamefont
  {Sch\"aefer}},\ }\href {\doibase 10.1103/PhysRevLett.66.2549} {\bibfield
  {journal} {\bibinfo  {journal} {Phys. Rev. Lett.}\ }\textbf {\bibinfo
  {volume} {66}},\ \bibinfo {pages} {2549} (\bibinfo {year}
  {1991})}\BibitemShut {NoStop}%
\bibitem [{\citenamefont {Freire}\ \emph {et~al.}(2012)\citenamefont {Freire},
  \citenamefont {Kramer},\ and\ \citenamefont {Wex}}]{Freire:2012nb}%
  \BibitemOpen
  \bibfield  {author} {\bibinfo {author} {\bibfnamefont {P.~C.~C.}\
  \bibnamefont {Freire}}, \bibinfo {author} {\bibfnamefont {M.}~\bibnamefont
  {Kramer}}, \ and\ \bibinfo {author} {\bibfnamefont {N.}~\bibnamefont {Wex}},\
  }\href {\doibase 10.1088/0264-9381/29/18/184007} {\bibfield  {journal}
  {\bibinfo  {journal} {Class. Quant. Grav.}\ }\textbf {\bibinfo {volume}
  {29}},\ \bibinfo {pages} {184007} (\bibinfo {year} {2012})},\ \Eprint
  {http://arxiv.org/abs/1205.3751} {arXiv:1205.3751 [gr-qc]} \BibitemShut
  {NoStop}%
\bibitem [{\citenamefont {Archibald}\ \emph {et~al.}(2018)\citenamefont
  {Archibald}, \citenamefont {Gusinskaia}, \citenamefont {Hessels},
  \citenamefont {Deller}, \citenamefont {Kaplan}, \citenamefont {Lorimer},
  \citenamefont {Lynch}, \citenamefont {Ransom},\ and\ \citenamefont
  {Stairs}}]{Archibald:2018oxs}%
  \BibitemOpen
  \bibfield  {author} {\bibinfo {author} {\bibfnamefont {A.~M.}\ \bibnamefont
  {Archibald}}, \bibinfo {author} {\bibfnamefont {N.~V.}\ \bibnamefont
  {Gusinskaia}}, \bibinfo {author} {\bibfnamefont {J.~W.~T.}\ \bibnamefont
  {Hessels}}, \bibinfo {author} {\bibfnamefont {A.~T.}\ \bibnamefont {Deller}},
  \bibinfo {author} {\bibfnamefont {D.~L.}\ \bibnamefont {Kaplan}}, \bibinfo
  {author} {\bibfnamefont {D.~R.}\ \bibnamefont {Lorimer}}, \bibinfo {author}
  {\bibfnamefont {R.~S.}\ \bibnamefont {Lynch}}, \bibinfo {author}
  {\bibfnamefont {S.~M.}\ \bibnamefont {Ransom}}, \ and\ \bibinfo {author}
  {\bibfnamefont {I.~H.}\ \bibnamefont {Stairs}},\ }\href {\doibase
  10.1038/s41586-018-0265-1} {\bibfield  {journal} {\bibinfo  {journal}
  {Nature}\ }\textbf {\bibinfo {volume} {559}},\ \bibinfo {pages} {73}
  (\bibinfo {year} {2018})},\ \Eprint {http://arxiv.org/abs/1807.02059}
  {arXiv:1807.02059 [astro-ph.HE]} \BibitemShut {NoStop}%
\bibitem [{\citenamefont {Stairs}(2003)}]{Stairs:2003eg}%
  \BibitemOpen
  \bibfield  {author} {\bibinfo {author} {\bibfnamefont {I.~H.}\ \bibnamefont
  {Stairs}},\ }\href {\doibase 10.12942/lrr-2003-5} {\bibfield  {journal}
  {\bibinfo  {journal} {Living Rev. Rel.}\ }\textbf {\bibinfo {volume} {6}},\
  \bibinfo {pages} {5} (\bibinfo {year} {2003})},\ \Eprint
  {http://arxiv.org/abs/astro-ph/0307536} {arXiv:astro-ph/0307536 [astro-ph]}
  \BibitemShut {NoStop}%
\bibitem [{\citenamefont {Hou}\ and\ \citenamefont {Gong}(2018)}]{Hou:2017cjy}%
  \BibitemOpen
  \bibfield  {author} {\bibinfo {author} {\bibfnamefont {S.}~\bibnamefont
  {Hou}}\ and\ \bibinfo {author} {\bibfnamefont {Y.}~\bibnamefont {Gong}},\
  }\href {\doibase 10.1140/epjc/s10052-018-5738-8} {\bibfield  {journal}
  {\bibinfo  {journal} {Eur. Phys. J. C}\ }\textbf {\bibinfo {volume} {78}},\
  \bibinfo {pages} {247} (\bibinfo {year} {2018})},\ \Eprint
  {http://arxiv.org/abs/1711.05034} {arXiv:1711.05034 [gr-qc]} \BibitemShut
  {NoStop}%
\bibitem [{\citenamefont {Zhu}\ \emph {et~al.}(2018)\citenamefont {Zhu} \emph
  {et~al.}}]{Zhu:2018etc}%
  \BibitemOpen
  \bibfield  {author} {\bibinfo {author} {\bibfnamefont {W.~W.}\ \bibnamefont
  {Zhu}} \emph {et~al.},\ }\href {\doibase 10.1093/mnras/sty2905} {\bibfield
  {journal} {\bibinfo  {journal} {Mon. Not. R. Astron. Soc.}\ }\textbf
  {\bibinfo {volume} {482}},\ \bibinfo {pages} {3249} (\bibinfo {year}
  {2018})},\ \Eprint {http://arxiv.org/abs/1802.09206} {arXiv:1802.09206
  [astro-ph.HE]} \BibitemShut {NoStop}%
\bibitem [{\citenamefont {Jenet}\ \emph {et~al.}(2005)\citenamefont {Jenet},
  \citenamefont {Hobbs}, \citenamefont {Lee},\ and\ \citenamefont
  {Manchester}}]{Jenet:2005pv}%
  \BibitemOpen
  \bibfield  {author} {\bibinfo {author} {\bibfnamefont {F.~A.}\ \bibnamefont
  {Jenet}}, \bibinfo {author} {\bibfnamefont {G.~B.}\ \bibnamefont {Hobbs}},
  \bibinfo {author} {\bibfnamefont {K.~J.}\ \bibnamefont {Lee}}, \ and\
  \bibinfo {author} {\bibfnamefont {R.~N.}\ \bibnamefont {Manchester}},\ }\href
  {\doibase 10.1086/431220} {\bibfield  {journal} {\bibinfo  {journal}
  {Astrophys. J.}\ }\textbf {\bibinfo {volume} {625}},\ \bibinfo {pages} {L123}
  (\bibinfo {year} {2005})},\ \Eprint {http://arxiv.org/abs/astro-ph/0504458}
  {arXiv:astro-ph/0504458 [astro-ph]} \BibitemShut {NoStop}%
\bibitem [{\citenamefont {{Lee}}\ \emph {et~al.}(2008)\citenamefont {{Lee}},
  \citenamefont {{Jenet}},\ and\ \citenamefont
  {{Price}}}]{2008ApJ...685.1304L}%
  \BibitemOpen
  \bibfield  {author} {\bibinfo {author} {\bibfnamefont {K.~J.}\ \bibnamefont
  {{Lee}}}, \bibinfo {author} {\bibfnamefont {F.~A.}\ \bibnamefont {{Jenet}}},
  \ and\ \bibinfo {author} {\bibfnamefont {R.~H.}\ \bibnamefont {{Price}}},\
  }\href {\doibase 10.1086/591080} {\bibfield  {journal} {\bibinfo  {journal}
  {Astrophys. J.}\ }\textbf {\bibinfo {volume} {685}},\ \bibinfo {eid}
  {1304-1319} (\bibinfo {year} {2008})}\BibitemShut {NoStop}%
\bibitem [{\citenamefont {Lee}\ \emph {et~al.}(2010)\citenamefont {Lee},
  \citenamefont {Jenet}, \citenamefont {Price}, \citenamefont {Wex},\ and\
  \citenamefont {Kramer}}]{Lee:2010cg}%
  \BibitemOpen
  \bibfield  {author} {\bibinfo {author} {\bibfnamefont {K.}~\bibnamefont
  {Lee}}, \bibinfo {author} {\bibfnamefont {F.~A.}\ \bibnamefont {Jenet}},
  \bibinfo {author} {\bibfnamefont {R.~H.}\ \bibnamefont {Price}}, \bibinfo
  {author} {\bibfnamefont {N.}~\bibnamefont {Wex}}, \ and\ \bibinfo {author}
  {\bibfnamefont {M.}~\bibnamefont {Kramer}},\ }\href {\doibase
  10.1088/0004-637X/722/2/1589} {\bibfield  {journal} {\bibinfo  {journal}
  {Astrophys. J.}\ }\textbf {\bibinfo {volume} {722}},\ \bibinfo {pages} {1589}
  (\bibinfo {year} {2010})},\ \Eprint {http://arxiv.org/abs/1008.2561}
  {arXiv:1008.2561 [astro-ph.HE]} \BibitemShut {NoStop}%
\bibitem [{\citenamefont {{Lee}}(2013)}]{Lee:2014awa}%
  \BibitemOpen
  \bibfield  {author} {\bibinfo {author} {\bibfnamefont {K.~J.}\ \bibnamefont
  {{Lee}}},\ }\href {\doibase 10.1088/0264-9381/30/22/224016} {\bibfield
  {journal} {\bibinfo  {journal} {Class. Quant. Grav.}\ }\textbf {\bibinfo
  {volume} {30}},\ \bibinfo {eid} {224016} (\bibinfo {year}
  {2013})}\BibitemShut {NoStop}%
\bibitem [{\citenamefont {Penrose}\ and\ \citenamefont
  {Rindler}(1984)}]{Penrose1984v1}%
  \BibitemOpen
  \bibfield  {author} {\bibinfo {author} {\bibfnamefont {R.}~\bibnamefont
  {Penrose}}\ and\ \bibinfo {author} {\bibfnamefont {W.}~\bibnamefont
  {Rindler}},\ }\href@noop {} {\emph {\bibinfo {title} {Spinors and
  space-time}}},\ \bibinfo {edition} {1st}\ ed.,\ \bibinfo {series} {Cambridge
  Monographs on Mathematical Physics}, Vol.~\bibinfo {volume} {1}\ (\bibinfo
  {publisher} {Cambridge University Press},\ \bibinfo {year}
  {1984})\BibitemShut {NoStop}%
\bibitem [{\citenamefont {Wald}(1984)}]{Wald:1984rg}%
  \BibitemOpen
  \bibfield  {author} {\bibinfo {author} {\bibfnamefont {R.~M.}\ \bibnamefont
  {Wald}},\ }\href {\doibase 10.7208/chicago/9780226870373.001.0001} {\emph
  {\bibinfo {title} {{General Relativity}}}}\ (\bibinfo  {publisher}
  {University of Chicago Press},\ \bibinfo {address} {Chicago, IL},\ \bibinfo
  {year} {1984})\BibitemShut {NoStop}%
\bibitem [{\citenamefont {{Swaminarayan}}\ and\ \citenamefont
  {{Safko}}(1983)}]{1983JMP....24..883S}%
  \BibitemOpen
  \bibfield  {author} {\bibinfo {author} {\bibfnamefont {N.~S.}\ \bibnamefont
  {{Swaminarayan}}}\ and\ \bibinfo {author} {\bibfnamefont {J.~L.}\
  \bibnamefont {{Safko}}},\ }\href {\doibase 10.1063/1.525776} {\bibfield
  {journal} {\bibinfo  {journal} {J. Math. Phys.}\ }\textbf {\bibinfo {volume}
  {24}},\ \bibinfo {pages} {883} (\bibinfo {year} {1983})}\BibitemShut
  {NoStop}%
\bibitem [{\citenamefont {Misner}\ \emph {et~al.}(1973)\citenamefont {Misner},
  \citenamefont {Thorne},\ and\ \citenamefont {Wheeler}}]{mtw}%
  \BibitemOpen
  \bibfield  {author} {\bibinfo {author} {\bibfnamefont {C.~W.}\ \bibnamefont
  {Misner}}, \bibinfo {author} {\bibfnamefont {K.~S.}\ \bibnamefont {Thorne}},
  \ and\ \bibinfo {author} {\bibfnamefont {J.~A.}\ \bibnamefont {Wheeler}},\
  }\href@noop {} {\emph {\bibinfo {title} {Gravitation}}}\ (\bibinfo
  {publisher} {San Francisco, W.H. Freeman},\ \bibinfo {year}
  {1973})\BibitemShut {NoStop}%
\bibitem [{\citenamefont {Hawking}\ and\ \citenamefont
  {Ellis}(2011)}]{Hawking:1973uf}%
  \BibitemOpen
  \bibfield  {author} {\bibinfo {author} {\bibfnamefont {S.~W.}\ \bibnamefont
  {Hawking}}\ and\ \bibinfo {author} {\bibfnamefont {G.~F.~R.}\ \bibnamefont
  {Ellis}},\ }\href {\doibase 10.1017/CBO9780511524646} {\emph {\bibinfo
  {title} {{The Large Scale Structure of Space-Time}}}},\ Cambridge Monographs
  on Mathematical Physics\ (\bibinfo  {publisher} {Cambridge University
  Press},\ \bibinfo {year} {2011})\BibitemShut {NoStop}%
\bibitem [{\citenamefont {Kobayashi}\ \emph {et~al.}(2011)\citenamefont
  {Kobayashi}, \citenamefont {Yamaguchi},\ and\ \citenamefont
  {Yokoyama}}]{Kobayashi:2011nu}%
  \BibitemOpen
  \bibfield  {author} {\bibinfo {author} {\bibfnamefont {T.}~\bibnamefont
  {Kobayashi}}, \bibinfo {author} {\bibfnamefont {M.}~\bibnamefont
  {Yamaguchi}}, \ and\ \bibinfo {author} {\bibfnamefont {J.}~\bibnamefont
  {Yokoyama}},\ }\href {\doibase 10.1143/PTP.126.511} {\bibfield  {journal}
  {\bibinfo  {journal} {Prog. Theor. Phys.}\ }\textbf {\bibinfo {volume}
  {126}},\ \bibinfo {pages} {511} (\bibinfo {year} {2011})},\ \Eprint
  {http://arxiv.org/abs/1105.5723} {arXiv:1105.5723 [hep-th]} \BibitemShut
  {NoStop}%
\bibitem [{\citenamefont {Gao}(2011)}]{Gao:2011mz}%
  \BibitemOpen
  \bibfield  {author} {\bibinfo {author} {\bibfnamefont {X.}~\bibnamefont
  {Gao}},\ }\href {\doibase 10.1088/1475-7516/2011/10/021} {\bibfield
  {journal} {\bibinfo  {journal} {JCAP}\ }\textbf {\bibinfo {volume} {1110}},\
  \bibinfo {pages} {021} (\bibinfo {year} {2011})},\ \Eprint
  {http://arxiv.org/abs/1106.0292} {arXiv:1106.0292 [astro-ph.CO]} \BibitemShut
  {NoStop}%
\bibitem [{\citenamefont {Abbott}\ \emph
  {et~al.}(2017{\natexlab{e}})\citenamefont {Abbott} \emph
  {et~al.}}]{Monitor:2017mdv}%
  \BibitemOpen
  \bibfield  {author} {\bibinfo {author} {\bibfnamefont {B.~P.}\ \bibnamefont
  {Abbott}} \emph {et~al.} (\bibinfo {collaboration} {Virgo, Fermi-GBM,
  INTEGRAL, LIGO Scientific}),\ }\href {\doibase 10.3847/2041-8213/aa920c}
  {\bibfield  {journal} {\bibinfo  {journal} {Astrophys. J.}\ }\textbf
  {\bibinfo {volume} {848}},\ \bibinfo {pages} {L13} (\bibinfo {year}
  {2017}{\natexlab{e}})},\ \Eprint {http://arxiv.org/abs/1710.05834}
  {arXiv:1710.05834 [astro-ph.HE]} \BibitemShut {NoStop}%
\bibitem [{\citenamefont {Lombriser}\ and\ \citenamefont
  {Taylor}(2016)}]{Lombriser:2015sxa}%
  \BibitemOpen
  \bibfield  {author} {\bibinfo {author} {\bibfnamefont {L.}~\bibnamefont
  {Lombriser}}\ and\ \bibinfo {author} {\bibfnamefont {A.}~\bibnamefont
  {Taylor}},\ }\href {\doibase 10.1088/1475-7516/2016/03/031} {\bibfield
  {journal} {\bibinfo  {journal} {JCAP}\ }\textbf {\bibinfo {volume} {1603}},\
  \bibinfo {pages} {031} (\bibinfo {year} {2016})},\ \Eprint
  {http://arxiv.org/abs/1509.08458} {arXiv:1509.08458 [astro-ph.CO]}
  \BibitemShut {NoStop}%
\bibitem [{\citenamefont {Lombriser}\ and\ \citenamefont
  {Lima}(2017)}]{Lombriser:2016yzn}%
  \BibitemOpen
  \bibfield  {author} {\bibinfo {author} {\bibfnamefont {L.}~\bibnamefont
  {Lombriser}}\ and\ \bibinfo {author} {\bibfnamefont {N.~A.}\ \bibnamefont
  {Lima}},\ }\href {\doibase 10.1016/j.physletb.2016.12.048} {\bibfield
  {journal} {\bibinfo  {journal} {Phys. Lett. B}\ }\textbf {\bibinfo {volume}
  {765}},\ \bibinfo {pages} {382} (\bibinfo {year} {2017})},\ \Eprint
  {http://arxiv.org/abs/1602.07670} {arXiv:1602.07670 [astro-ph.CO]}
  \BibitemShut {NoStop}%
\bibitem [{\citenamefont {Baker}\ \emph {et~al.}(2017)\citenamefont {Baker},
  \citenamefont {Bellini}, \citenamefont {Ferreira}, \citenamefont {Lagos},
  \citenamefont {Noller},\ and\ \citenamefont {Sawicki}}]{Baker:2017hug}%
  \BibitemOpen
  \bibfield  {author} {\bibinfo {author} {\bibfnamefont {T.}~\bibnamefont
  {Baker}}, \bibinfo {author} {\bibfnamefont {E.}~\bibnamefont {Bellini}},
  \bibinfo {author} {\bibfnamefont {P.~G.}\ \bibnamefont {Ferreira}}, \bibinfo
  {author} {\bibfnamefont {M.}~\bibnamefont {Lagos}}, \bibinfo {author}
  {\bibfnamefont {J.}~\bibnamefont {Noller}}, \ and\ \bibinfo {author}
  {\bibfnamefont {I.}~\bibnamefont {Sawicki}},\ }\href {\doibase
  10.1103/PhysRevLett.119.251301} {\bibfield  {journal} {\bibinfo  {journal}
  {Phys. Rev. Lett.}\ }\textbf {\bibinfo {volume} {119}},\ \bibinfo {pages}
  {251301} (\bibinfo {year} {2017})},\ \Eprint
  {http://arxiv.org/abs/1710.06394} {arXiv:1710.06394 [astro-ph.CO]}
  \BibitemShut {NoStop}%
\bibitem [{\citenamefont {Creminelli}\ and\ \citenamefont
  {Vernizzi}(2017)}]{Creminelli:2017sry}%
  \BibitemOpen
  \bibfield  {author} {\bibinfo {author} {\bibfnamefont {P.}~\bibnamefont
  {Creminelli}}\ and\ \bibinfo {author} {\bibfnamefont {F.}~\bibnamefont
  {Vernizzi}},\ }\href {\doibase 10.1103/PhysRevLett.119.251302} {\bibfield
  {journal} {\bibinfo  {journal} {Phys. Rev. Lett.}\ }\textbf {\bibinfo
  {volume} {119}},\ \bibinfo {pages} {251302} (\bibinfo {year} {2017})},\
  \Eprint {http://arxiv.org/abs/1710.05877} {arXiv:1710.05877 [astro-ph.CO]}
  \BibitemShut {NoStop}%
\bibitem [{\citenamefont {Sakstein}\ and\ \citenamefont
  {Jain}(2017)}]{Sakstein:2017xjx}%
  \BibitemOpen
  \bibfield  {author} {\bibinfo {author} {\bibfnamefont {J.}~\bibnamefont
  {Sakstein}}\ and\ \bibinfo {author} {\bibfnamefont {B.}~\bibnamefont
  {Jain}},\ }\href {\doibase 10.1103/PhysRevLett.119.251303} {\bibfield
  {journal} {\bibinfo  {journal} {Phys. Rev. Lett.}\ }\textbf {\bibinfo
  {volume} {119}},\ \bibinfo {pages} {251303} (\bibinfo {year} {2017})},\
  \Eprint {http://arxiv.org/abs/1710.05893} {arXiv:1710.05893 [astro-ph.CO]}
  \BibitemShut {NoStop}%
\bibitem [{\citenamefont {Ezquiaga}\ and\ \citenamefont
  {Zumalac\'arregui}(2017)}]{Ezquiaga:2017ekz}%
  \BibitemOpen
  \bibfield  {author} {\bibinfo {author} {\bibfnamefont {J.~M.}\ \bibnamefont
  {Ezquiaga}}\ and\ \bibinfo {author} {\bibfnamefont {M.}~\bibnamefont
  {Zumalac\'arregui}},\ }\href {\doibase 10.1103/PhysRevLett.119.251304}
  {\bibfield  {journal} {\bibinfo  {journal} {Phys. Rev. Lett.}\ }\textbf
  {\bibinfo {volume} {119}},\ \bibinfo {pages} {251304} (\bibinfo {year}
  {2017})},\ \Eprint {http://arxiv.org/abs/1710.05901} {arXiv:1710.05901
  [astro-ph.CO]} \BibitemShut {NoStop}%
\bibitem [{\citenamefont {Langlois}\ \emph {et~al.}(2018)\citenamefont
  {Langlois}, \citenamefont {Saito}, \citenamefont {Yamauchi},\ and\
  \citenamefont {Noui}}]{Langlois:2017dyl}%
  \BibitemOpen
  \bibfield  {author} {\bibinfo {author} {\bibfnamefont {D.}~\bibnamefont
  {Langlois}}, \bibinfo {author} {\bibfnamefont {R.}~\bibnamefont {Saito}},
  \bibinfo {author} {\bibfnamefont {D.}~\bibnamefont {Yamauchi}}, \ and\
  \bibinfo {author} {\bibfnamefont {K.}~\bibnamefont {Noui}},\ }\href {\doibase
  10.1103/PhysRevD.97.061501} {\bibfield  {journal} {\bibinfo  {journal} {Phys.
  Rev. D}\ }\textbf {\bibinfo {volume} {97}},\ \bibinfo {pages} {061501}
  (\bibinfo {year} {2018})},\ \Eprint {http://arxiv.org/abs/1711.07403}
  {arXiv:1711.07403 [gr-qc]} \BibitemShut {NoStop}%
\bibitem [{\citenamefont {Gong}\ \emph {et~al.}(2018)\citenamefont {Gong},
  \citenamefont {Papantonopoulos},\ and\ \citenamefont {Yi}}]{Gong:2017kim}%
  \BibitemOpen
  \bibfield  {author} {\bibinfo {author} {\bibfnamefont {Y.}~\bibnamefont
  {Gong}}, \bibinfo {author} {\bibfnamefont {E.}~\bibnamefont
  {Papantonopoulos}}, \ and\ \bibinfo {author} {\bibfnamefont {Z.}~\bibnamefont
  {Yi}},\ }\href {\doibase 10.1140/epjc/s10052-018-6227-9} {\bibfield
  {journal} {\bibinfo  {journal} {Eur. Phys. J. C}\ }\textbf {\bibinfo {volume}
  {78}},\ \bibinfo {pages} {738} (\bibinfo {year} {2018})},\ \Eprint
  {http://arxiv.org/abs/1711.04102} {arXiv:1711.04102 [gr-qc]} \BibitemShut
  {NoStop}%
\bibitem [{\citenamefont {Hohmann}(2015)}]{Hohmann:2015kra}%
  \BibitemOpen
  \bibfield  {author} {\bibinfo {author} {\bibfnamefont {M.}~\bibnamefont
  {Hohmann}},\ }\href {\doibase 10.1103/PhysRevD.92.064019} {\bibfield
  {journal} {\bibinfo  {journal} {Phys. Rev. D}\ }\textbf {\bibinfo {volume}
  {92}},\ \bibinfo {pages} {064019} (\bibinfo {year} {2015})},\ \Eprint
  {http://arxiv.org/abs/1506.04253} {arXiv:1506.04253 [gr-qc]} \BibitemShut
  {NoStop}%
\bibitem [{\citenamefont {{Eardley}}(1975)}]{1975ApJ...196L..59E}%
  \BibitemOpen
  \bibfield  {author} {\bibinfo {author} {\bibfnamefont {D.~M.}\ \bibnamefont
  {{Eardley}}},\ }\href {\doibase 10.1086/181744} {\bibfield  {journal}
  {\bibinfo  {journal} {Astrophys. J.}\ }\textbf {\bibinfo {volume} {196}},\
  \bibinfo {pages} {L59} (\bibinfo {year} {1975})}\BibitemShut {NoStop}%
\bibitem [{\citenamefont {Hou}\ \emph {et~al.}(2018)\citenamefont {Hou},
  \citenamefont {Gong},\ and\ \citenamefont {Liu}}]{Hou:2017bqj}%
  \BibitemOpen
  \bibfield  {author} {\bibinfo {author} {\bibfnamefont {S.}~\bibnamefont
  {Hou}}, \bibinfo {author} {\bibfnamefont {Y.}~\bibnamefont {Gong}}, \ and\
  \bibinfo {author} {\bibfnamefont {Y.}~\bibnamefont {Liu}},\ }\href {\doibase
  10.1140/epjc/s10052-018-5869-y} {\bibfield  {journal} {\bibinfo  {journal}
  {Eur. Phys. J. C}\ }\textbf {\bibinfo {volume} {78}},\ \bibinfo {pages} {378}
  (\bibinfo {year} {2018})},\ \Eprint {http://arxiv.org/abs/1704.01899}
  {arXiv:1704.01899 [gr-qc]} \BibitemShut {NoStop}%
\bibitem [{\citenamefont {{Fortini}}\ and\ \citenamefont
  {{Gualdi}}(1982)}]{1982NCimB..71...37F}%
  \BibitemOpen
  \bibfield  {author} {\bibinfo {author} {\bibfnamefont {P.~L.}\ \bibnamefont
  {{Fortini}}}\ and\ \bibinfo {author} {\bibfnamefont {C.}~\bibnamefont
  {{Gualdi}}},\ }\href {\doibase 10.1007/BF02721692} {\bibfield  {journal}
  {\bibinfo  {journal} {Nuovo Cimento B}\ }\textbf {\bibinfo {volume} {71}},\
  \bibinfo {pages} {37} (\bibinfo {year} {1982})}\BibitemShut {NoStop}%
\bibitem [{\citenamefont {{Fortini}}\ and\ \citenamefont
  {{Ortolan}}(1991)}]{1991NCimB.106..101F}%
  \BibitemOpen
  \bibfield  {author} {\bibinfo {author} {\bibfnamefont {P.}~\bibnamefont
  {{Fortini}}}\ and\ \bibinfo {author} {\bibfnamefont {A.}~\bibnamefont
  {{Ortolan}}},\ }\href {\doibase 10.1007/BF02723131} {\bibfield  {journal}
  {\bibinfo  {journal} {Nuovo Cimento B}\ }\textbf {\bibinfo {volume} {106}},\
  \bibinfo {pages} {101} (\bibinfo {year} {1991})}\BibitemShut {NoStop}%
\bibitem [{\citenamefont {Zaglauer}(1992)}]{Zaglauer1992}%
  \BibitemOpen
  \bibfield  {author} {\bibinfo {author} {\bibfnamefont {H.~W.}\ \bibnamefont
  {Zaglauer}},\ }\href {\doibase 10.1086/171537} {\bibfield  {journal}
  {\bibinfo  {journal} {Astrophys. J.}\ }\textbf {\bibinfo {volume} {393}},\
  \bibinfo {pages} {685} (\bibinfo {year} {1992})}\BibitemShut {NoStop}%
\bibitem [{\citenamefont {Berti}\ \emph {et~al.}(2015)\citenamefont {Berti}
  \emph {et~al.}}]{Berti:2015itd}%
  \BibitemOpen
  \bibfield  {author} {\bibinfo {author} {\bibfnamefont {E.}~\bibnamefont
  {Berti}} \emph {et~al.},\ }\href {\doibase 10.1088/0264-9381/32/24/243001}
  {\bibfield  {journal} {\bibinfo  {journal} {Class. Quant. Grav.}\ }\textbf
  {\bibinfo {volume} {32}},\ \bibinfo {pages} {243001} (\bibinfo {year}
  {2015})},\ \Eprint {http://arxiv.org/abs/1501.07274} {arXiv:1501.07274
  [gr-qc]} \BibitemShut {NoStop}%
\bibitem [{\citenamefont {{Lorimer}}\ and\ \citenamefont
  {{Kramer}}(2004)}]{2004hpa..book.....L}%
  \BibitemOpen
  \bibfield  {author} {\bibinfo {author} {\bibfnamefont {D.~R.}\ \bibnamefont
  {{Lorimer}}}\ and\ \bibinfo {author} {\bibfnamefont {M.}~\bibnamefont
  {{Kramer}}},\ }\href@noop {} {\emph {\bibinfo {title} {Handbook of pulsar
  astronomy, by D.R.~Lorimer and M.~Kramer.~Cambridge observing handbooks for
  research astronomers, Vol.~4.~Cambridge, UK: Cambridge University Press,
  2004}}}\ (\bibinfo {year} {2004})\BibitemShut {NoStop}%
\bibitem [{\citenamefont {Verbiest}\ \emph {et~al.}(2009)\citenamefont
  {Verbiest} \emph {et~al.}}]{Verbiest:2009kb}%
  \BibitemOpen
  \bibfield  {author} {\bibinfo {author} {\bibfnamefont {J.~P.~W.}\
  \bibnamefont {Verbiest}} \emph {et~al.},\ }\href {\doibase
  10.1111/j.1365-2966.2009.15508.x} {\bibfield  {journal} {\bibinfo  {journal}
  {Mon. Not. Roy. Astron. Soc.}\ }\textbf {\bibinfo {volume} {400}},\ \bibinfo
  {pages} {951} (\bibinfo {year} {2009})},\ \Eprint
  {http://arxiv.org/abs/0908.0244} {arXiv:0908.0244 [astro-ph.GA]} \BibitemShut
  {NoStop}%
\bibitem [{\citenamefont {{Estabrook}}\ and\ \citenamefont
  {{Wahlquist}}(1975)}]{1975GReGr...6..439E}%
  \BibitemOpen
  \bibfield  {author} {\bibinfo {author} {\bibfnamefont {F.~B.}\ \bibnamefont
  {{Estabrook}}}\ and\ \bibinfo {author} {\bibfnamefont {H.~D.}\ \bibnamefont
  {{Wahlquist}}},\ }\href {\doibase 10.1007/BF00762449} {\bibfield  {journal}
  {\bibinfo  {journal} {Gen. Rel. Grav.}\ }\textbf {\bibinfo {volume} {6}},\
  \bibinfo {pages} {439} (\bibinfo {year} {1975})}\BibitemShut {NoStop}%
\bibitem [{\citenamefont {Sazhin}(1978)}]{1978SvA....22...36S}%
  \BibitemOpen
  \bibfield  {author} {\bibinfo {author} {\bibfnamefont {M.~V.}\ \bibnamefont
  {Sazhin}},\ }\href {http://adsabs.harvard.edu/abs/1978SvA....22...36S}
  {\bibfield  {journal} {\bibinfo  {journal} {Sov. Astron.}\ }\textbf {\bibinfo
  {volume} {22}},\ \bibinfo {pages} {36} (\bibinfo {year} {1978})}\BibitemShut
  {NoStop}%
\bibitem [{\citenamefont {Detweiler}(1979)}]{Detweiler:1979wn}%
  \BibitemOpen
  \bibfield  {author} {\bibinfo {author} {\bibfnamefont {S.~L.}\ \bibnamefont
  {Detweiler}},\ }\href {\doibase 10.1086/157593} {\bibfield  {journal}
  {\bibinfo  {journal} {Astrophys. J.}\ }\textbf {\bibinfo {volume} {234}},\
  \bibinfo {pages} {1100} (\bibinfo {year} {1979})}\BibitemShut {NoStop}%
\bibitem [{\citenamefont {Hellings}\ and\ \citenamefont
  {Downs}(1983)}]{Hellings:1983fr}%
  \BibitemOpen
  \bibfield  {author} {\bibinfo {author} {\bibfnamefont {R.~W.}\ \bibnamefont
  {Hellings}}\ and\ \bibinfo {author} {\bibfnamefont {G.~S.}\ \bibnamefont
  {Downs}},\ }\href {\doibase 10.1086/183954} {\bibfield  {journal} {\bibinfo
  {journal} {Astrophys. J.}\ }\textbf {\bibinfo {volume} {265}},\ \bibinfo
  {pages} {L39} (\bibinfo {year} {1983})}\BibitemShut {NoStop}%
\bibitem [{\citenamefont {Chamberlin}\ and\ \citenamefont
  {Siemens}(2012)}]{Chamberlin:2011ev}%
  \BibitemOpen
  \bibfield  {author} {\bibinfo {author} {\bibfnamefont {S.~J.}\ \bibnamefont
  {Chamberlin}}\ and\ \bibinfo {author} {\bibfnamefont {X.}~\bibnamefont
  {Siemens}},\ }\href {\doibase 10.1103/PhysRevD.85.082001} {\bibfield
  {journal} {\bibinfo  {journal} {Phys. Rev. D}\ }\textbf {\bibinfo {volume}
  {85}},\ \bibinfo {pages} {082001} (\bibinfo {year} {2012})},\ \Eprint
  {http://arxiv.org/abs/1111.5661} {arXiv:1111.5661 [astro-ph.HE]} \BibitemShut
  {NoStop}%
\bibitem [{\citenamefont {Yunes}\ and\ \citenamefont
  {Siemens}(2013)}]{Yunes:2013dva}%
  \BibitemOpen
  \bibfield  {author} {\bibinfo {author} {\bibfnamefont {N.}~\bibnamefont
  {Yunes}}\ and\ \bibinfo {author} {\bibfnamefont {X.}~\bibnamefont
  {Siemens}},\ }\href {\doibase 10.12942/lrr-2013-9} {\bibfield  {journal}
  {\bibinfo  {journal} {Living Rev. Rel.}\ }\textbf {\bibinfo {volume} {16}},\
  \bibinfo {pages} {9} (\bibinfo {year} {2013})},\ \Eprint
  {http://arxiv.org/abs/1304.3473} {arXiv:1304.3473 [gr-qc]} \BibitemShut
  {NoStop}%
\bibitem [{\citenamefont {Gair}\ \emph {et~al.}(2014)\citenamefont {Gair},
  \citenamefont {Romano}, \citenamefont {Taylor},\ and\ \citenamefont
  {Mingarelli}}]{Gair:2014rwa}%
  \BibitemOpen
  \bibfield  {author} {\bibinfo {author} {\bibfnamefont {J.}~\bibnamefont
  {Gair}}, \bibinfo {author} {\bibfnamefont {J.~D.}\ \bibnamefont {Romano}},
  \bibinfo {author} {\bibfnamefont {S.}~\bibnamefont {Taylor}}, \ and\ \bibinfo
  {author} {\bibfnamefont {C.~M.~F.}\ \bibnamefont {Mingarelli}},\ }\href
  {\doibase 10.1103/PhysRevD.90.082001} {\bibfield  {journal} {\bibinfo
  {journal} {Phys. Rev. D}\ }\textbf {\bibinfo {volume} {90}},\ \bibinfo
  {pages} {082001} (\bibinfo {year} {2014})},\ \Eprint
  {http://arxiv.org/abs/1406.4664} {arXiv:1406.4664 [gr-qc]} \BibitemShut
  {NoStop}%
\bibitem [{\citenamefont {Gair}\ \emph {et~al.}(2015)\citenamefont {Gair},
  \citenamefont {Romano},\ and\ \citenamefont {Taylor}}]{Gair:2015hra}%
  \BibitemOpen
  \bibfield  {author} {\bibinfo {author} {\bibfnamefont {J.~R.}\ \bibnamefont
  {Gair}}, \bibinfo {author} {\bibfnamefont {J.~D.}\ \bibnamefont {Romano}}, \
  and\ \bibinfo {author} {\bibfnamefont {S.~R.}\ \bibnamefont {Taylor}},\
  }\href {\doibase 10.1103/PhysRevD.92.102003} {\bibfield  {journal} {\bibinfo
  {journal} {Phys. Rev. D}\ }\textbf {\bibinfo {volume} {92}},\ \bibinfo
  {pages} {102003} (\bibinfo {year} {2015})},\ \Eprint
  {http://arxiv.org/abs/1506.08668} {arXiv:1506.08668 [gr-qc]} \BibitemShut
  {NoStop}%
\bibitem [{\citenamefont {Romano}\ and\ \citenamefont
  {Cornish}(2017)}]{Romano:2016dpx}%
  \BibitemOpen
  \bibfield  {author} {\bibinfo {author} {\bibfnamefont {J.~D.}\ \bibnamefont
  {Romano}}\ and\ \bibinfo {author} {\bibfnamefont {N.~J.}\ \bibnamefont
  {Cornish}},\ }\href {\doibase 10.1007/s41114-017-0004-1} {\bibfield
  {journal} {\bibinfo  {journal} {Living Rev. Rel.}\ }\textbf {\bibinfo
  {volume} {20}},\ \bibinfo {pages} {2} (\bibinfo {year} {2017})},\ \Eprint
  {http://arxiv.org/abs/1608.06889} {arXiv:1608.06889 [gr-qc]} \BibitemShut
  {NoStop}%
\bibitem [{\citenamefont {Arzoumanian}\ \emph {et~al.}(2018)\citenamefont
  {Arzoumanian} \emph {et~al.}}]{Arzoumanian:2018saf}%
  \BibitemOpen
  \bibfield  {author} {\bibinfo {author} {\bibfnamefont {Z.}~\bibnamefont
  {Arzoumanian}} \emph {et~al.} (\bibinfo {collaboration} {NANOGRAV}),\ }\href
  {\doibase 10.3847/1538-4357/aabd3b} {\bibfield  {journal} {\bibinfo
  {journal} {Astrophys. J.}\ }\textbf {\bibinfo {volume} {859}},\ \bibinfo
  {pages} {47} (\bibinfo {year} {2018})},\ \Eprint
  {http://arxiv.org/abs/1801.02617} {arXiv:1801.02617 [astro-ph.HE]}
  \BibitemShut {NoStop}%
\bibitem [{Note2()}]{Note2}%
  \BibitemOpen
  \bibinfo {note} {Precisely due to the large errors, we do not plot the
  observed cross correlation in our Fig.~\ref {fig-grbrsen}.}\BibitemShut
  {Stop}%
\bibitem [{\citenamefont {Liang}\ \emph {et~al.}(2017)\citenamefont {Liang},
  \citenamefont {Gong}, \citenamefont {Hou},\ and\ \citenamefont
  {Liu}}]{Liang:2017ahj}%
  \BibitemOpen
  \bibfield  {author} {\bibinfo {author} {\bibfnamefont {D.}~\bibnamefont
  {Liang}}, \bibinfo {author} {\bibfnamefont {Y.}~\bibnamefont {Gong}},
  \bibinfo {author} {\bibfnamefont {S.}~\bibnamefont {Hou}}, \ and\ \bibinfo
  {author} {\bibfnamefont {Y.}~\bibnamefont {Liu}},\ }\href {\doibase
  10.1103/PhysRevD.95.104034} {\bibfield  {journal} {\bibinfo  {journal} {Phys.
  Rev. D}\ }\textbf {\bibinfo {volume} {95}},\ \bibinfo {pages} {104034}
  (\bibinfo {year} {2017})},\ \Eprint {http://arxiv.org/abs/1701.05998}
  {arXiv:1701.05998 [gr-qc]} \BibitemShut {NoStop}%
\bibitem [{\citenamefont {Gong}\ and\ \citenamefont
  {Hou}(2018{\natexlab{a}})}]{Gong:2017bru}%
  \BibitemOpen
  \bibfield  {author} {\bibinfo {author} {\bibfnamefont {Y.}~\bibnamefont
  {Gong}}\ and\ \bibinfo {author} {\bibfnamefont {S.}~\bibnamefont {Hou}},\
  }\bibfield  {booktitle} {\emph {\bibinfo {booktitle} {{Proceedings, 13th
  International Conference on Gravitation, Astrophysics and Cosmology and 15th
  Italian-Korean Symposium on Relativistic Astrophysics (IK15): Seoul, Korea,
  July 3-7, 2017}}},\ }\href {\doibase 10.1051/epjconf/201816801003} {\bibfield
   {journal} {\bibinfo  {journal} {EPJ Web Conf.}\ }\textbf {\bibinfo {volume}
  {168}},\ \bibinfo {pages} {01003} (\bibinfo {year} {2018}{\natexlab{a}})},\
  \Eprint {http://arxiv.org/abs/1709.03313} {arXiv:1709.03313 [gr-qc]}
  \BibitemShut {NoStop}%
\bibitem [{\citenamefont {Gong}\ and\ \citenamefont
  {Hou}(2018{\natexlab{b}})}]{Gong:2018ybk}%
  \BibitemOpen
  \bibfield  {author} {\bibinfo {author} {\bibfnamefont {Y.}~\bibnamefont
  {Gong}}\ and\ \bibinfo {author} {\bibfnamefont {S.}~\bibnamefont {Hou}},\
  }\bibfield  {booktitle} {\emph {\bibinfo {booktitle} {{International
  Conference on Quantum Gravity Shenzhen, Guangdong, China, March 26-28,
  2018}}},\ }\href {\doibase 10.3390/universe4080085} {\  (\bibinfo {year}
  {2018}{\natexlab{b}}),\ 10.3390/universe4080085},\ \bibinfo {note}
  {[Universe4,no.8,85(2018)]},\ \Eprint {http://arxiv.org/abs/1806.04027}
  {arXiv:1806.04027 [gr-qc]} \BibitemShut {NoStop}%
\bibitem [{\citenamefont {Deruelle}(2011)}]{Deruelle:2011wu}%
  \BibitemOpen
  \bibfield  {author} {\bibinfo {author} {\bibfnamefont {N.}~\bibnamefont
  {Deruelle}},\ }\href {\doibase 10.1007/s10714-011-1247-x} {\bibfield
  {journal} {\bibinfo  {journal} {Gen. Rel. Grav.}\ }\textbf {\bibinfo {volume}
  {43}},\ \bibinfo {pages} {3337} (\bibinfo {year} {2011})},\ \Eprint
  {http://arxiv.org/abs/1104.4608} {arXiv:1104.4608 [gr-qc]} \BibitemShut
  {NoStop}%
\end{thebibliography}

%

\end{document}